\documentclass[10pt,english]{article}
\usepackage{newcent}
\usepackage[T1]{fontenc}
\usepackage[latin1]{inputenc}
\usepackage{geometry}
\usepackage{graphicx}
\IfFileExists{url.sty}{\usepackage{url}}
                      {\newcommand{\url}{\texttt}}

\makeatletter

\providecommand{\tabularnewline}{\\}

 \newcommand{\lyxaddress}[1]{
   \par {\raggedright #1 
   \vspace{1.4em}
   \noindent\par}
 }

\usepackage{hyperref}
\hypersetup{colorlinks=true,bookmarks=true,bookmarksnumbered=true,pdftitle="Technique(s) For Spike-Sorting",pdfauthor="Christophe Pouzat"}

\AtBeginDocument{
  
}

\usepackage{babel}
\makeatother
\begin{document}

\title{TECHNIQUE(S) FOR SPIKE - SORTING}

\author{Christophe Pouzat}

\maketitle

\lyxaddress{Laboratoire de Physiologie Cérébrale, CNRS UMR 8118\\
UFR Biomédicale de l'Université René Descartes (Paris V)\\
45, rue des Saints Pères\\
75006 Paris\\
France\\
\url{http://www.biomedicale.univ-paris5.fr/physcerv/Spike-O-Matic.html}\\
\url{e-mail: christophe.pouzat@univ-paris5.fr}}

\lyxaddress{\pagebreak}

\tableofcontents{}

\pagebreak

\section{Introduction}

Most of this book is dedicated to the presentation of models of neuronal
networks and of methods developed to work with them. Within the frameworks
set by these models, the activity of populations of neurons is derived
based on the intrinsic properties of {}``simplified'' neurons and
of their connectivity pattern. The experimentalist trying to test
these models must start by collecting data for which model predictions
can be made, which means he or she must record the activity of several
neurons at once. If the model does in fact predict the compound activity
of a local neuronal population (\emph{e.g.}, a cortical column or
hyper-column) on a relatively slow time scale (100 msec or more),
techniques like intrinsic optical imaging \cite{FrostigEtAl1990}
are perfectly adapted. If the model makes {}``more precise'' predictions
on the activity of individual neurons on a short time scale (\textasciitilde{}
msec) then appropriate techniques must be used like fast optical recording
with voltage sensitive dyes \cite{WuEtAl1994} or the more classical
extracellular recording technique \cite{Lewicki1998}.

We will focus in this chapter on the problems associated with the
extracellular recording technique which is still a very popular investigation
method. This popularity is partly due to its relative ease of implementation
and to its low cost. The author is moreover clearly biased toward
this technique being one of its users. We hope nevertheless that users
of other techniques will learn something from what follows. We will
explain how to make inferences about values of the parameters of a
rather complex model from noisy data. The approach we will develop
can be adapted to any problem of this type. What will change from
an experimental context to another is the relevant data generation
model and the noise model. That does not mean that the adaptation
of the method to other contexts is trivial, but it is still doable.

The data generation model with which we will work in this chapter
is qualitatively different from the ones which have been considered
in the past \cite{Lewicki1994,Sahani1999,PouzatEtAl2002,NugyenEtAl2003}.
The reader unfamilliar with the spike-sorting problem and willing
to get some background on it can consult with profit Lewicki's review
\cite{Lewicki1998}. To get an introduction on the actual statistical
methods adapted to the models previously considered, we recommend
the consultation of the statistical literature (\emph{eg}, \cite{CeleuxGovaert1995})
rather than the spike-sorting one, which tends, in our opinion, to
suffer from a strong taste for ad-hoc methods.

\section{The problem to solve}

\begin{figure}
\begin{center}\includegraphics[%
  scale=0.7]{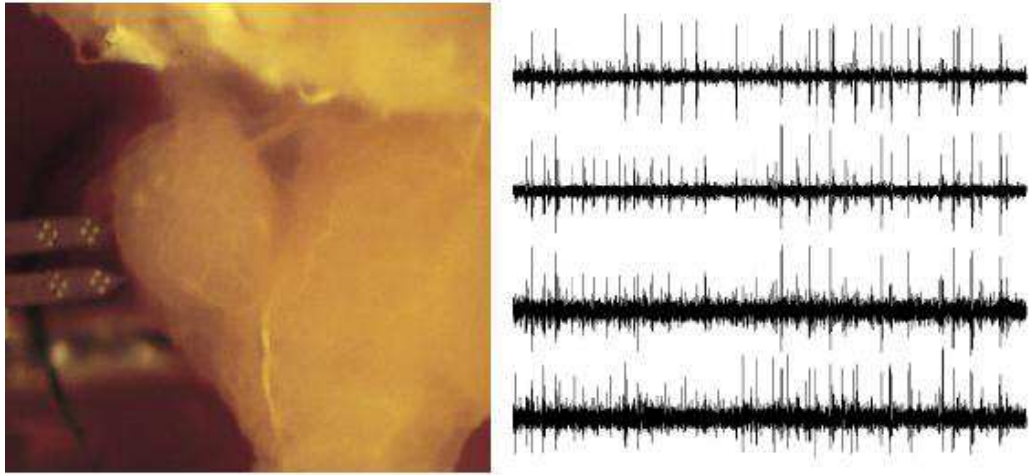}\end{center}

\caption{\label{cap:Example raw data 1 s}Example of a tetrode recording from
the locust (\emph{Schistocerca americana}) antennal lobe. Left, recording
setting. The probe, a silicon substrate made of two shanks with 16
iridium deposits (the actual recording sites), can be seen on the
left side of the picture. Each group of 4 recording sites is called
a \emph{tetrode} (there are therefore 4 tetrodes by probe). The recording
sites are the bright spots. The width of the shanks is 80 $\mu m$
, the side length of each recording site is 13 $\mu m$ , the diagonal
length of each tetrode is 50 $\mu m$ , the center to center distance
between neighboring tetrodes is 150 $\mu m$ . The structure right
beside the probe tip is the \emph{antennal lobe} (the first olfactory
relay of the insect), its diameter is approximately 400 $\mu m$ .
Once the probe has been gently pushed into the antennal lobe such
that the lowest two tetrodes are roughly 100 $\mu m$ below the surface
one gets on these lowest tetrodes data looking typically as shown
on the right part of the figure. Right, 1s of data from a single tetrode
filtered between 300 and 5kHz.}
\end{figure}

\begin{figure}
\begin{center}\includegraphics[%
  scale=0.4]{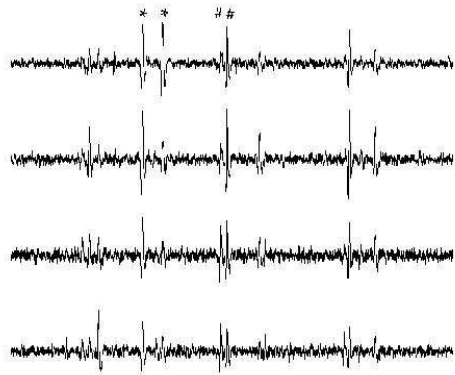}\end{center}

\caption{\label{cap:Example raw data 200 ms}The last 200 ms of Fig. \ref{cap:Example raw data 1 s}.
Considering two pairs of spikes ({*}{*} and \#\#) the interest of
the tetrodes becomes clear. On the first recording site (top) the
two spikes of the pair ({*}{*}) look very similar and it would therefore
be hard for an analyst having only the top recording site information
to say if these two spikes originate from the same neuron or from
two different ones. If now, one looks at the same spikes on the three
other recording sites the difference is obvious. The same holds for
the two spikes of the pair (\#\#). They look similar on sites 3 and
4 but very dissimilar on sites 1 and 2.}
\end{figure}

Extracellular recordings are typically a mixture of spike waveforms
originating from a generally unknown number of neurons to which a
background noise is superposed as illustrated on Fig \ref{cap:Example raw data 1 s}.%
 Several features can be used to distinguish spikes from different
neurons \cite{Lewicki1998} like the peak amplitude on a single of
several recording sites (Fig \ref{cap:Example raw data 200 ms}),
the spike width, a bi - or tri - phasic waveform, etc. 

\subsection*{What do we want?}

\begin{itemize}
\item Find the number of neurons contributing to the data.
\item Find the value of a set of parameters characterizing the signal generated
by each neuron (\emph{e.g.}, the spike waveform of each neuron on
each recording site).
\item Acknowledging the classification ambiguity which can arise from waveform
similarity and/or signal corruption due to noise, the probability
for each neuron to have generated each event (spike) in the data set.
\item A method as automatic as possible.
\item A non ad - hoc method to answer the above questions. By non ad - hoc
we mean a method based on an \emph{explicit} \emph{probabilistic}
model for data generation.
\end{itemize}

\section{Two features of single neuron data we would like to include in the
spike-sorting procedure}

\subsection{Spike waveforms \emph{from a single neuron} are usually not stationary
on a short time - scale}

\subsubsection{An experimental illustration with cerebellar Purkinje cells}

\begin{figure}
\begin{center}\includegraphics[%
  scale=0.5]{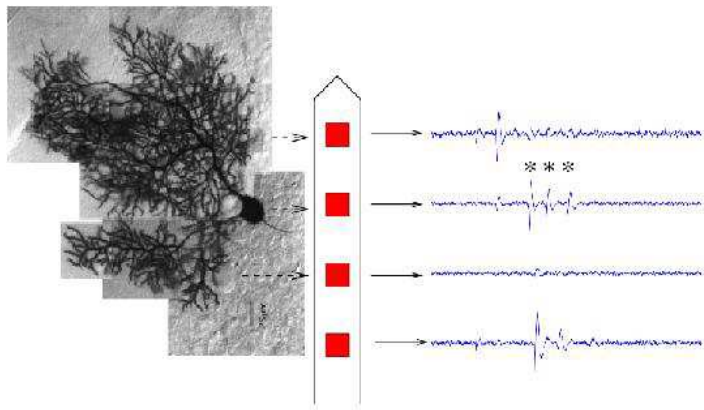}\end{center}

\caption{An example of a multi - electrode recording from a rat cerebellar
slice. Left, picture of a slice (from a 38 days old rat) with a biocytin
filled Purkinje cell. Middle, the tip of a Michigan probe drawn to
scale. Right, 100 ms of data filtered between 300 and 5K Hz, Scale
bar: 10 ms, Stars: a triplet of spike coming from a single Purkinje
cell. (Delescluse and Pouzat, unpublished) \label{cap:Fig Amp vs ISI}}
\end{figure}

One commonly observes that, when principal cells fire bursts of action
potentials, the spike amplitude decreases%
\footnote{More generally the spike shape changes and basically slows down \cite{HarrisEtAl2000}.
This is mainly due to sodium channels inactivation.%
} during the burst as illustrated on Fig. \ref{cap:Fig Amp vs ISI}.

\subsubsection{A phenomenological description by an exponential relaxation}

Following \cite{FeeEtAl1996} we will use an exponential relaxation
to describe the spike waveform dependence upon the inter - spike interval:\begin{equation}
\textbf{a}\left(isi\right)=\textbf{{p}}\cdot\left(1-\delta\cdot\exp\left(-\lambda\cdot isi\right)\right),\label{eq:Amp vs ISI}\end{equation}

where $\textbf{{p}}$ is the vector of maximal amplitudes (or full
waveforms) on the different recording sites, $\delta\in\left[0,1\right]$,
$\lambda$, measured in 1/s, is the inverse of the relaxation time
constant.

\subsection{Neurons inter - spike interval probability density functions carry
a lot of information we would like to exploit}

\subsubsection{An experimental illustration from projection neurons in the locust
antennal lobe}

\begin{figure}
\begin{center}\includegraphics[%
  scale=0.5]{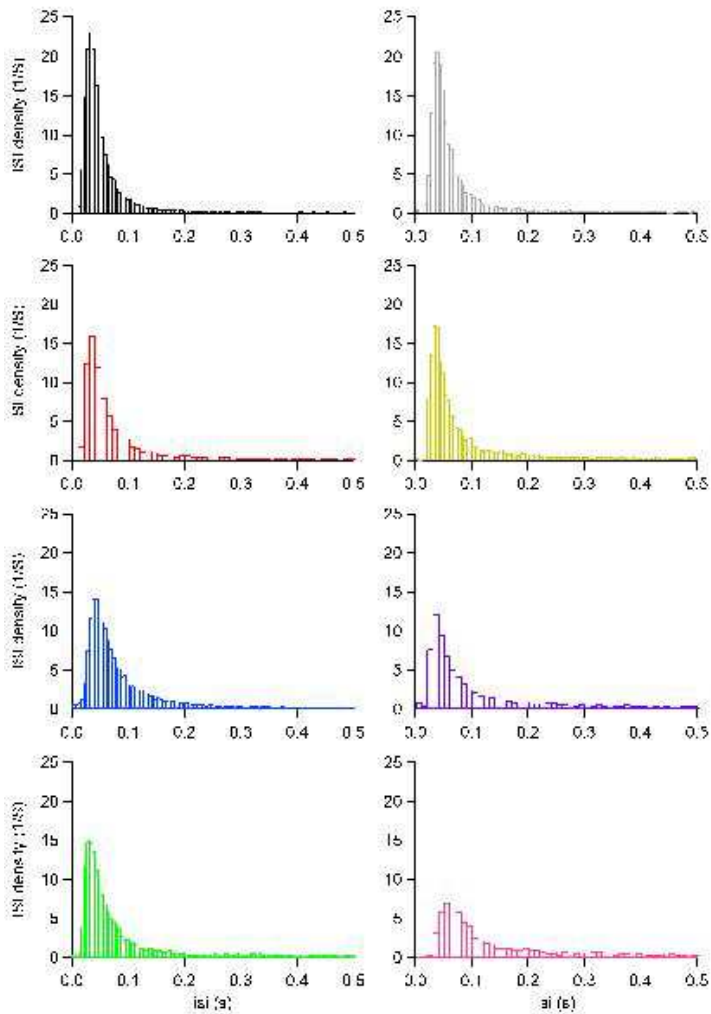}\end{center}

\caption{\label{cap:ISI PDF example}An example of \emph{ISI} \emph{pdf} estimates
for 8 projection neurons simultaneously recorded in the locust antennal
lobe (Pouzat, Mazor and Laurent, unpublished).}
\end{figure}

It is well known and understood since the squid giant axon study by
Hodgkin and Huxley that once a neuron (or a piece of it like its axon)
has fired an action potential, we must wait {}``a while'' before
the next action potential can be fired. This delay is dubbed the \emph{refractory
period} and is mainly due to inactivation of sodium channels and to
strong activation of potassium channels at the end of the spike, meaning
that we must wait for the potassium channels to de - activate and
for sodium channel to de - inactivate. Phenomenologically it means
that we should observe on the inter - spike interval (\emph{ISI})
histogram from a single neuron a period without spikes (\emph{i.e.},
the \emph{ISI} histogram should start at zero and stay at zero for
a finite time). In addition we can often find on \emph{ISI} histograms
some other features like a single mode%
\footnote{That's the statistician's terminology for local maximum.%
} a {}``fast'' rise and a {}``slower'' decay as illustrated on
Fig \ref{cap:ISI PDF example}. The knowledge of the \emph{ISI} histogram
can in principle be used to improve spike - sorting because it will
induce correlations between the labeling of successive spikes. Indeed,
if in one way or another we can be sure of the labeling of a given
spike to a given neuron, we can be sure as well that the probability
to have an other spike from the same neuron within say the next 10
ms is zero, that this probability is high between 30 and 50 ms and
high as well between 60 and 100 (you just need to integrate the \emph{ISI}
histogram to see that).

\subsubsection{A phenomenological description by a log - Normal \emph{pdf}}

\begin{figure}
\begin{center}\includegraphics[%
  scale=0.5]{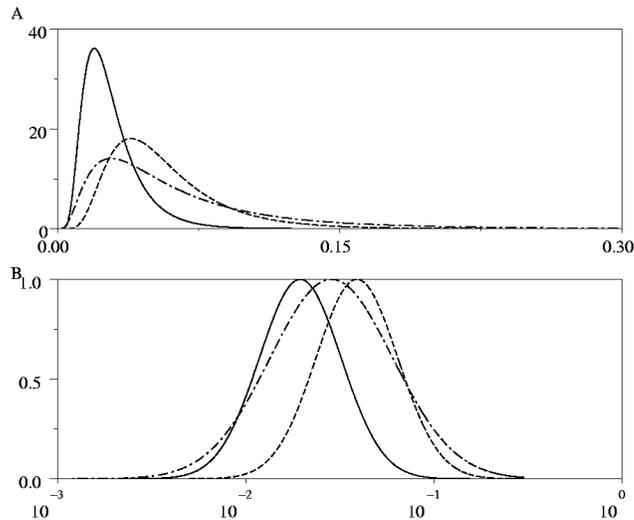}\end{center}

\caption{\label{cap:Examples Log-Normal}A, examples of log - Normal densities.
plain: \emph{S} = 0.025, \emph{F} = 0.5; dashed: \emph{S} = 0.05,
\emph{F} = 0.5; dot-dashed: \emph{S} = 0.05, \emph{F} = 0.75. B, peak
normalized densities displayed with a logarithmic abscissa.}
\end{figure}

We will use a log - Normal approximation for our empirical \emph{ISI}
probability density functions (\emph{pdf}s):\begin{equation}
\pi\left(ISI=isi\mid S=s,F=f\right)=\frac{1}{isi\cdot f\cdot\sqrt{2\pi}}\cdot\exp\left[-\frac{1}{2}\cdot\left(\frac{\ln isi-\ln s}{f}\right)^{2}\right],\label{eq:ISI PDF}\end{equation}

where S is a scale parameter (measured in seconds, like the \emph{ISI})
and $F$ is a shape parameter (dimensionless). Fig. \ref{cap:Examples Log-Normal}
shows three log - Normal densities for different values of \emph{S}
and \emph{F}. The lowest part of the figure shows that when we look
at the density of the logarithm of the \emph{ISI} we get a Normal
distribution which explains the name of this density.

In the sequel, we will in general use $\pi$ to designate proper probability
density functions (\emph{pdf} for continuous random variables) or
probability mass functions (\emph{pmf} for discrete random variables).
By proper we mean the integral (or the sum) over the appropriate space
is 1. We will (try to) use uppercases to designate random variables
(\emph{e.g.}, \emph{ISI, S,} $F$) and lowercases to designate their
realizations (\emph{e.g.}, \emph{isi}, \emph{s}, \emph{f}). We will
use a Bayesian approach, which means that the model parameters (like
\emph{S} and $F$ in Eq. \ref{eq:ISI PDF}) will be considered as
random variables.

\section{Noise properties}

\begin{figure}
\begin{center}\includegraphics[%
  scale=0.5]{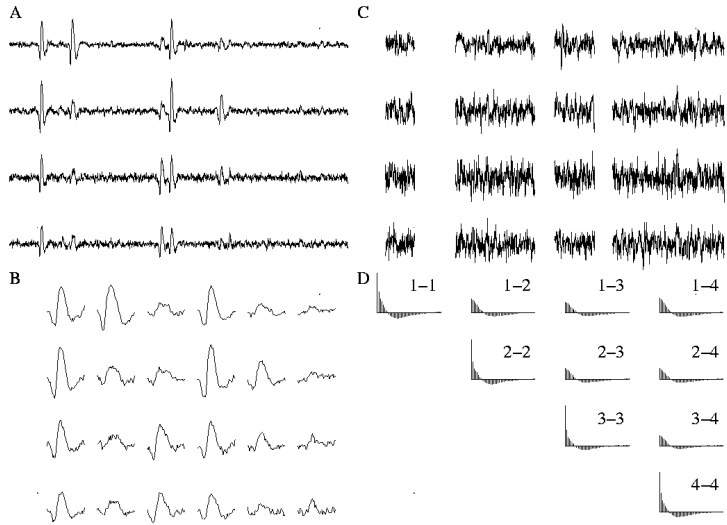}\end{center}

\caption{\label{cap:Noise Extraction}Procedure used to obtain the second
order statistical properties of the recording noise illustrated with
a tetrode recording from the locust antennal lobe. Putative {}``events''
(B, the sweeps are 3 ms long) and {}``noise'' (C) are separated
from the complete data (A) after a threshold crossing and template
matching detection of the events. The noise auto - and cross - correlation
functions (D, 3 ms are shown for each function, the autocorrelations
functions are on the diagonal, the crosscorrelation functions on the
upper part) are computed from the reconstructed noise traces (C).
Adapted from \cite{PouzatEtAl2002}.}
\end{figure}

It's better to start with a reasonably accurate noise model and Fig.
\ref{cap:Noise Extraction} illustrates how the noise statistical
properties can be obtained from actual data.%
 Once the empirical noise auto - and cross - correlation functions
have been obtained, assuming the noise is stationary, the noise covariance
matrix ($\Sigma$) is constructed as a block - Toeplitz matrix. There
are as many blocks as the square of the number of recording sites.
The first row of each block is the corresponding auto- or cross -
correlation function, see \cite{PouzatEtAl2002}. Then, if $\Sigma$
is a complete noise description, the \emph{pdf} of a noise vector
$\textbf{{N}}$ is multivariate Gaussian:\begin{equation}
\pi\left(\textbf{{N}}=\textbf{{n}}\right)=\frac{1}{\left(2\pi\right)^{\frac{D}{2}}}\cdot\left|\Sigma^{-1}\right|^{\frac{1}{2}}\cdot\exp\left(-\frac{1}{2}\cdot\textbf{{n}}^{T}\Sigma^{-1}\,\textbf{{n}}\right),\label{eq:Noise pdf}\end{equation}

where \emph{D} is the dimension of the space used to represent the
events (and the noise), $\left|\Sigma^{-1}\right|$ stands for the
determinant of the inverse of $\Sigma$, $\textbf{{n}}^{T}$ stands
for the transpose of $\textbf{{n}}$.

\subsection{Noise whitening}

If the noise covariance matrix is known, it is useful to know about
a transformation, \emph{noise whitening}, which is easy to perform.
It makes the equations easier to write and computations faster. If
$\Sigma$ is indeed a covariance matrix, that implies:\begin{equation}
\textbf{{v}}^{T}\Sigma^{-1}\,\textbf{{v}}\geq0,\forall\:\textbf{{v}}\,\in\Re^{D}.\label{eq:Proper Distance}\end{equation}

That is, the \emph{Mahalanobis distance} ($\textbf{{v}}^{T}\Sigma^{-1}\,\textbf{{v}}$)
is a proper distance which is stated differently by saying that the
inverse of $\Sigma$ is positive definite. Then one can show (Exercise
1 below) that there exists a unique lower triangular matrix \emph{A}
such that $AA^{T}=\Sigma^{-1}$ and transform vector $\textbf{{v}}$
of Eq. \ref{eq:Proper Distance} into $\textbf{{w}}=A^{T}\textbf{{v}}$.
We then have:\begin{equation}
\textbf{{w}}^{T}\textbf{{w}}=\textbf{{v}}^{T}\Sigma^{-1}\,\textbf{{v}}.\label{eq:Mahalanobis to Euclidian}\end{equation}

That is, we found a new coordinate system in which the \emph{Mahalanobis
distance} is the \emph{Euclidean distance}. We say we perform \emph{noise
whitening} when we go from $\textbf{{v}}$ to $\textbf{{w}}$. 

\begin{figure}
\begin{center}\includegraphics[%
  scale=0.4]{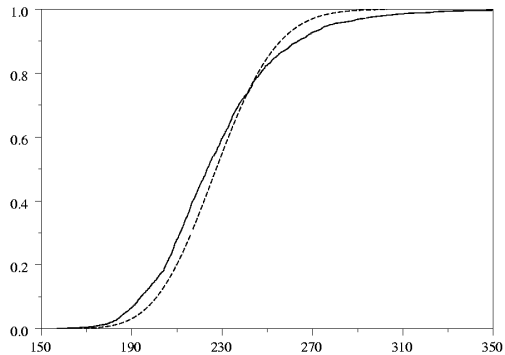}\end{center}

\caption{\label{cap:Noise Check}How good is a noise description based on
a multivariate Gaussian \emph{pdf}? \emph{}A typical example (from
locust antennal lobe) where $\Sigma$ turns out to be a reasonable
noise description. A sample of 2000 noise events $\left\{ \textbf{{n}}_{1},\ldots,\textbf{{n}}_{2000}\right\} $
was randomly selected from the reconstructed noise trace (Fig. \ref{cap:Noise Extraction}C),
the \emph{Mahalanobis distance} to the origin: $\textbf{{n}}_{i}^{T}\Sigma^{-1}\,\textbf{{n}}_{i}$,
was computed for each of them and the cumulative distribution of these
distances (plain) was compared with the theoretical expectation (dashed)
a $\chi^{2}$distribution (with in that case 238 degrees of freedom).
See \cite{PouzatEtAl2002} for details.}
\end{figure}

Assuming the noise is stationary and fully described by its second
order statistical properties is not enough. One can check the validity
of this assumption after whitening an empirical noise sample as illustrated
in Fig. \ref{cap:Noise Check}.

\subsubsection*{Exercise 1}

Assume that $\Gamma=\Sigma^{-1}$ is a symmetric positive definite
matrix. Show there exists a unique (in fact 2, one being obtained
from the other by a multiplication by - 1) lower triangular matrix
\emph{A} (the \emph{Cholesky} factor\emph{)} such that:\begin{equation}
AA^{T}=\Gamma.\label{eq:Cholesky Decomposition}\end{equation}

In order to do that you will simply get the algorithm which computes
the elements of \emph{A} from the elements of $\Gamma$. See Sec.
\ref{sub:Cholesky-decomposition} for solution.

\vspace{0.2cm}

\section{Probabilistic data generation model}

\subsection{Model assumptions}

We will make the following assumptions:

\begin{enumerate}
\item The firing statistics of each neuron is fully described by its time
independent inter - spike interval density. That is, the sequence
of spike times from a given neuron is a realization of \emph{a homogeneous
renewal point process} \cite{Johnson1996}. More specifically, we
will assume that the \emph{ISI pdf} is log - Normal (Eq. \ref{eq:ISI PDF}).
\item The amplitude of the spikes generated by each neuron depends on the
elapsed time since the last spike of this neuron. More specifically,
we will assume that this dependence is well described by Eq. \ref{eq:Amp vs ISI}.
\item The measured amplitude of the spikes is corrupted by a Gaussian white
noise which sums linearly with the spikes and is statistically independent
of them. That is, we assume that noise whitening (Eq. \ref{eq:Mahalanobis to Euclidian})
has been performed.
\end{enumerate}

\subsection{Likelihood computation for single neuron data\label{sub:Single-Neuron-Likelihood-Comp}}

\begin{figure}
\begin{center}\includegraphics[%
  scale=0.5]{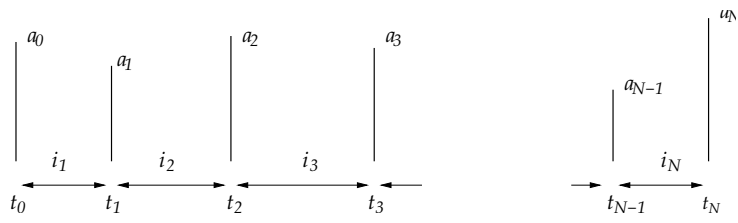}\end{center}

\caption{\label{cap:Single Neuron Likelohood}Illustration of the likelihood
computation for data from a single neuron recorded on a single site.
For simplicity we assume that the spike peak amplitude only is used.}
\end{figure}

With a data set $\mathcal{{D}}'=\left\{ \left(t_{0},\textbf{{a}}_{0}\right),\left(t_{1},\textbf{{a}}_{1}\right),\ldots,\left(t_{N},\textbf{{a}}_{N}\right)\right\} $,
the likelihood is readily obtained (Fig. \ref{cap:Single Neuron Likelohood}
). One must first get the \emph{isi} values: $i_{j}=t_{j}-t_{j-1}$,
which implies one has to discard the first spike of the data set%
\footnote{To avoid discarding the first spike we can assume periodic boundary
conditions, meaning the last spike (\emph{N}) {}``precedes'' the
first one (\emph{0}), we then have: $i_{0}=T-t_{N}+t_{0}$ , where
\emph{T} is the duration of the recording which started at time 0. %
} to get the {}``effective'' data set: $\mathcal{{D}}=\left\{ \left(i_{1},\textbf{{a}}_{1}\right),\ldots,\left(i_{N},\textbf{{a}}_{N}\right)\right\} $.
The likelihood is then%
\footnote{Purists from both sides (frequentist and Bayesian) would kill us for
writing what follows (Eq. \ref{eq:Single Neuron Likelihood})... A
frequentist would rather write:\[
L\left(\textbf{{p}},\delta,\lambda,s,f\,;\mathcal{{D}}\,\right)\]

because for him the likelihood function is a random function (it depends
on the sample through $\mathcal{{D}}$ ) and its arguments are the
model parameters. The Bayesian, probably to show the frequentist that
he perfectly understood the meaning of the likelihood function, would
introduce a new symbol, say \emph{h} and write:\[
h\left(\mathcal{{D}}\,\mid\textbf{{p}},\delta,\lambda,s,f\,\right)=L\left(\textbf{{p}},\delta,\lambda,s,f\,;\mathcal{{D}}\,\right).\]

But the likelihood is, within a normalizing constant, the probability
of the data for given values of the model parameters. We will therefore
use the heretic notation of Eq. \ref{eq:Single Neuron Likelihood}.%
}:\begin{equation}
L\left(\mathcal{{D}}\,\mid\textbf{{p}},\delta,\lambda,s,f\,\right)=\prod_{j=1}^{N}\pi_{isi}\left(i_{j}\mid s,f\right)\cdot\pi_{amplitude}\left(\textbf{{a}}_{j}\mid i_{j},\textbf{{p}},\delta,\lambda\right),\label{eq:Single Neuron Likelihood}\end{equation}

where $\pi_{isi}\left(i_{j}\mid s,f\right)$ is given by Eq. \ref{eq:ISI PDF}
and:\begin{equation}
\pi_{amplitude}\left(\textbf{{a}}_{j}\mid i_{j},\textbf{{p}},\delta,\lambda\right)=\frac{1}{\left(2\pi\right)^{\frac{D}{2}}}\cdot\exp\left\{ -\frac{1}{2}\left\Vert \textbf{{a}}_{j}-\textbf{{p}}\cdot\left(1-\delta\cdot\exp\left(-\lambda\cdot i_{j}\right)\right)\right\Vert ^{2}\right\} .\label{eq:Amplitude given ISI}\end{equation}

This last equation is obtained by combining Eq. \ref{eq:Amp vs ISI}
with our third model assumption (Gaussian white noise). For convenience
we write $\pi_{isi}\left(i_{j}\mid s,f\right)$ for $\pi_{isi}\left(ISI_{j}=i_{j}\mid S=s,F=f\right)$
and \\
$\pi_{amplitude}\left(\textbf{{a}}_{j}\mid i_{j},\textbf{{p}},\delta,\lambda\right)$
for $\pi_{amplitude}\left(\textbf{{A}}_{j}=\textbf{{a}}_{j}\mid ISI_{j}=i_{j},\textbf{{P}}=\textbf{{p}},\Delta=\delta,\Lambda=\lambda\right)$,
where the $ISI_{j}$ are considered independent and identically distributed
(\emph{iid}) random variables (conditioned on \emph{S} and \emph{F})
as well as the $\textbf{{A}}_{j}$ (conditioned on $ISI_{j}$, $\textbf{{P}}$,
$\Delta$ and $\Lambda$). 

The log - likelihood ($\mathcal{{L}}=\ln L$) can be written as a
sum of two terms:\begin{equation}
\mathcal{{L}}\left(\mathcal{{D}}\,\mid\textbf{{p}},\delta,\lambda,s,f\,\right)=\mathcal{{L}}_{isi}\left(\mathcal{{D}}\,\mid s,f\,\right)+\mathcal{{L}}_{amplitude}\left(\mathcal{{D}}\,\mid\textbf{{p}},\delta,\lambda\,\right),\label{eq:Single Neuron Log-Likelihodd}\end{equation}

where:\begin{equation}
\mathcal{{L}}_{isi}\left(\mathcal{{D}}\,\mid s,f\,\right)=-N\cdot\ln f-\sum_{j=1}^{N}\left\{ \ln i_{j}+\frac{1}{2}\left[\frac{\ln\left(\frac{i_{j}}{s}\right)}{f}\right]^{2}\right\} +Cst,\label{eq:Single Neuron Log-Likelihood ISI}\end{equation}

and:\begin{equation}
\mathcal{{L}}_{amplitude}\left(\mathcal{{D}}\,\mid\textbf{{p}},\delta,\lambda\,\right)=-\frac{1}{2}\sum_{j=1}^{N}\left\Vert \textbf{{a}}_{j}-\textbf{{p}}\cdot\left(1-\delta\cdot\exp\left(-\lambda\cdot i_{j}\right)\right)\right\Vert ^{2}+Cst.\label{eq:Single Neuron Log-Likelihood Amplitude}\end{equation}

The term $\mathcal{{L}}_{isi}$ is a function of the \emph{ISI pdf}
parameters only and the term $\mathcal{{L}}_{amplitude}$ is a function
of the {}``amplitude dynamics parameters'' only.

\subsubsection*{Exercise 2}

Starting with the expression of the log - likelihood given by Eq.
\ref{eq:Single Neuron Log-Likelihood ISI} , show that the values
$\widetilde{s}$ and $\widetilde{f}$ of \emph{s} and \emph{f} maximizing
the likelihood are given by:\begin{equation}
\ln\widetilde{s}=\frac{1}{N}\sum_{j=1}^{N}\ln i_{j},\label{eq:Best Scale}\end{equation}

and:\begin{equation}
\widetilde{f}=\sqrt{\frac{1}{N}\sum_{j=1}^{N}\left[\ln\left(\frac{i_{j}}{\widetilde{s}}\right)\right]^{2}}.\label{eq:Best Shape}\end{equation}

\subsubsection*{Exercise 3\label{sub:Exercice-3}}

Remember that in a Bayesian setting your start with a prior density
for your model parameters: $\pi_{prior}\left(s,f\right)$, you observe
some data, $\mathcal{{D}}$ , you have an explicit data generation
model (\emph{i.e.}, you know how to write the likelihood function),
$L\left(\mathcal{{D}}\,\mid s,f\,\right)$ and you can therefore write
using Bayes rule:\begin{equation}
\pi\left(s,f\,\mid\mathcal{{D}}\right)=\frac{L\left(\mathcal{{D}}\,\mid s,f\,\right)\cdot\pi_{prior}\left(s,f\right)}{\int_{u,v}\, du\, dv\, L\left(\mathcal{{D}}\,\mid u,v\,\right)\cdot\pi_{prior}\left(u,v\right)}\label{eq:Bayes Rule 1}\end{equation}

The {}``Bayesian game'' in simple cases is always the same, you
take the numerator of Eq. \ref{eq:Bayes Rule 1}, you get rid of everything
which does not involve the model parameters (because that will be
absorbed in the normalizing constant) and you try to recognize in
what's left a known \emph{pdf} like a Normal, a Gamma, a Beta, etc.

In the following we will assume that our prior density is uniform
on a rectangle, that is:\begin{equation}
\pi_{prior}\left(s,f\right)=\frac{1}{s_{max}-s_{min}}\cdot\mathcal{{I}}_{\left[s_{min},s_{max}\right]}\left(s\right)\cdot\frac{1}{f_{max}-f_{min}}\cdot\mathcal{{I}}_{\left[f_{min},f_{max}\right]}\left(f\right),\label{eq:Scale Shape Prior}\end{equation}

where $\mathcal{{I}}_{\left[x,y\right]}$is the indicator function:\begin{equation}
\mathcal{{I}}_{\left[x,y\right]}\left(u\right)=\left\{ \begin{array}{c}
\:\:1,\, if\, x\leq u\leq y\\
0,\, otherwise\end{array}\right.\label{eq:Indicator Function}\end{equation}

Detailed answers to the following questions can be found in Sec. \ref{sub:Bayesian-for-log-Normal}.

\paragraph{Question 1}

Using Eq. \ref{cap:Single Neuron Likelohood} and Eq. \ref{eq:ISI PDF}
show that the posterior density of $\ln S$ conditioned on $F=f$
and $\mathcal{{D}}$ is a truncated Gaussian with mean: \begin{equation}
\overline{\ln i}=\frac{1}{N}\,\sum_{j=1}^{N}\ln i_{j},\label{eq:Mean Log ISI}\end{equation}

and un-truncated variance (that is, the variance it would have if
it was not truncated):\begin{equation}
\sigma^{2}=\frac{f^{2}}{N}.\label{eq:Posterior Scale Variance}\end{equation}

And therefore the posterior conditional \emph{pdf} of the scale parameter
is given by:\begin{equation}
\pi\left(s\,\mid f,\mathcal{{D}}\right)\,\alpha\,\exp\left[-\frac{1}{2}\,\frac{N}{f^{2}}\,\left(\overline{\ln i}-\ln s\right)^{2}\right]\cdot\frac{1}{s_{max}-s_{min}}\cdot\mathcal{{I}}_{\left[s_{min},s_{max}\right]}\left(s\right).\label{eq:Posterior Scale PDF}\end{equation}

\paragraph*{Question 2}

Write an algorithm which generates a realization of \emph{S} according
to Eq. \ref{eq:Posterior Scale PDF}.

\paragraph*{Question 3}

We say that a random variable $\Theta\:$ has an \emph{Inverse Gamma}
distribution and we write: $\Theta\,\sim\, Inv-Gamma\left(\alpha,\beta\right)\:$if
the \emph{pdf} of $\Theta$ is given by:\begin{equation}
\pi\left(\Theta=\theta\right)=\frac{\beta^{\alpha}}{\Gamma\left(\alpha\right)}\,\theta^{-\left(\alpha+1\right)}\,\exp\left(-\frac{\beta}{\theta}\right).\label{eq:Inverse Gamma PDF}\end{equation}

Using Eq. \ref{cap:Single Neuron Likelohood} and Eq. \ref{eq:ISI PDF}
show that the posterior density of $F^{2}$ conditioned on $S=s$
and $\mathcal{{D}}$ is a truncated \emph{Inverse Gamma} with parameters:\[
\alpha=\frac{N}{2}-1\]

and\[
\beta=\frac{1}{2}\,\sum_{j=1}^{N}\left(\ln i_{j}-\ln s\right)^{2}.\]

Therefore the posterior conditional \emph{pdf} of the shape parameter
is:\begin{equation}
\pi\left(f\,\mid s,\mathcal{{D}}\right)\,\alpha\,\frac{1}{f^{N}}\,\exp\left[-\frac{1}{2}\frac{\sum_{j=1}^{N}\left(\ln i_{j}-\ln s\right)^{2}}{f^{2}}\right]\cdot\frac{1}{f_{max}-f_{min}}\cdot\mathcal{{I}}_{\left[f_{min},f_{max}\right]}\left(f\right).\label{eq:Posterior Shape PDF}\end{equation}

\paragraph*{Question 4}

Assuming that your favorite analysis software (\emph{e.g.}, Scilab%
\footnote{\url{http://www.scilab.org}%
} or Matlab) or your favorite C library (\emph{e.g.}, GSL%
\footnote{The Gnu Scientific Library: \url{http://sources.redhat.com/gsl}%
}) has a \emph{Gamma} random number generator, write an algorithm which
generates a realization of \emph{F} from its posterior density Eq.
\ref{eq:Posterior Shape PDF}. The \emph{pdf} of a \emph{Gamma} random
variable $\Omega$ is given by:\begin{equation}
\pi\left(\Omega=\omega\right)=\frac{\beta^{\alpha}}{\Gamma\left(\alpha\right)}\,\omega^{\alpha-1}\,\exp\left(-\beta\cdot\omega\right).\label{eq:Gamma PDF}\end{equation}

\subsection{Complications with multi - neuron data}

\subsubsection{Notations for the multi - neuron model parameters}

In the following we will consider a model with \emph{K} different
neurons, which means we will have to estimate the value of a maximal
peak amplitude parameter ($\textbf{{P}}$), an attenuation parameter
($\Delta$), the inverse of a relaxation time constant ($\Lambda$),
a scale parameter (\emph{S}) and a shape parameter (\emph{F}) for
each of the \emph{K} neurons of our model. We will write $\Theta$
the random vector lumping all these individual parameters:\begin{equation}
\Theta=\left(\textbf{{P}}_{1},\Delta_{1},\Lambda_{1},S_{1},F_{1},\ldots,\textbf{{P}}_{K},\Delta_{K},\Lambda_{K},S_{K},F_{K}\right).\label{eq:Theta Definition}\end{equation}

\subsubsection{Configuration and \emph{data augmentation}}

When we deal with multi - neuron data we need to formalize our ignorance
about the origin of each individual spike. In order to do that, assuming
we work with a model with \emph{K} neurons, we will {}``attach''
to each spike, \emph{j}, in our data set ($\mathcal{{D}}$) a \emph{label},
$l_{j}\in\left\{ 1,\ldots,K\right\} $ whose value corresponds to
the number of the neuron which generated it. From this view point,
a spike - sorting procedure is a method to set the labels values.
But we don't know, at least until we have estimated the model parameters,
what is the value of each $l_{j}$. In such an uncertain situation
the best we can do is to consider $l_{j}$ as a realization of a random
variable $L_{j}$ taking values in $\left\{ 1,\ldots,K\right\} $,
then the best we can expect from a spike - sorting procedure is the
\emph{distribution} (or the \emph{pmf}) \emph{of} $L_{j}$ for $j\in\left\{ 1,\ldots,N\right\} $.

Our data generation model will necessarily induce a dependence among
the $L_{j}$ because we take into account both the \emph{ISI} densities
and the spike waveform dynamics, we will therefore introduce an other
random variable, the \emph{configuration}, \emph{C} defined as a vector
of $L_{j}$ that is: \begin{equation}
C=\left(L_{1},\ldots,L_{N}\right)^{T}.\label{eq:Configuration Definition}\end{equation}

The introduction of this random variable \emph{C} is a very common
procedure in the statistical literature where it is called \emph{data
augmentation}. The key idea being that we are given an \emph{incomplete}
data set ($\mathcal{{D}}$), which basically does not allow us to
write the likelihood easily, and if we knew some extra properties
of the data (like the \emph{configuration}), we could write down the
likelihood of the \emph{augmented} (or \emph{completed}) data in a
straightforward way%
\footnote{\emph{C} is also called a \emph{latent} variable in the statistical
literature. Statisticians, moreover, commonly use the symbol \emph{Z}
for our \emph{C}. We have change the convention for reasons which
will soon become clear.%
}. Indeed, with \emph{C} introduced the likelihood of the augmented
data ($L\left(\mathcal{{D}},c\mid\theta\right)$) is obtained in two
steps:

\begin{figure}
\begin{center}\includegraphics[%
  scale=0.5]{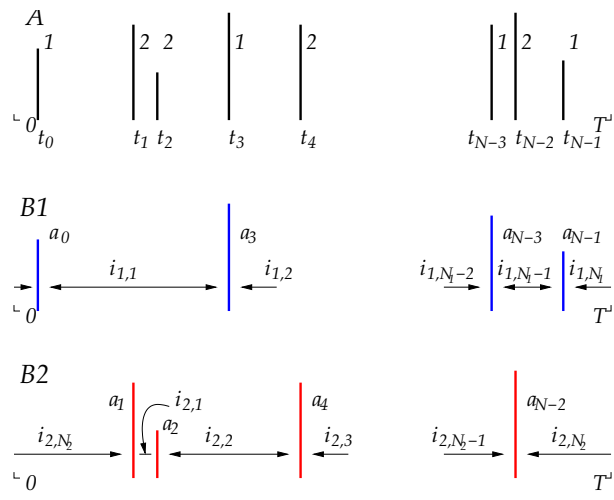}\end{center}

\caption{\label{cap:Multi Neuron Likelihood}A simple example of sub-trains
extraction. We consider a case with 2 neurons in the model and a single
recording site. A, Spikes are shown with their labels in the present
configuration and their occurrence times. The recording goes from
\emph{0} to \emph{T}. B1, the sub-train attributed to neuron 1 has
been isolated. The relevant \emph{ISIs} are shown, as well as the
amplitudes of the events. B2, same as B1 for the sub-train attributed
to neuron 2.}
\end{figure}

\begin{enumerate}
\item Using the configuration realization, \emph{c}, we can separate the
{}``complete'' data set into sub - trains corresponding to the different
neurons of the model. This is illustrated with a simple case, $K=2$
and a single recording site on Fig. \ref{cap:Multi Neuron Likelihood}.
Here we have: $i_{1,1}=t_{3}-t_{0}$ , $i_{1,N_{1}-1}=t_{N-3}-t_{N-1}$
, $i_{1,N_{1}}=T-t_{N-1}+t_{0}$, where periodic boundary conditions
are assumed and where $N_{1}$ is the number of spikes attributed
to neuron 1 in \emph{c}. For neuron 2 we get: $i_{2,1}=t_{2}-t_{1}$
, $i_{2,2}=t_{4}-t_{2}$ , $i_{2,N_{2}}=T-t_{N-2}+t_{1}$, where $N_{2}$
is the number of spikes attributed to neuron 2 in \emph{c}. Using
Eq. \ref{cap:Single Neuron Likelohood} we can compute a {}``sub
- likelihood'' for each sub - train: $L_{l_{j}=1}\left(\mathcal{{D}},c\mid\theta\right)$,
$L_{l_{j}=2}\left(\mathcal{{D}},c\mid\theta\right)$ .
\item Because our data generation model does not include interactions between
neurons, the {}``complete'' likelihood is simply%
\footnote{Strictly speaking, this is true only if we can ignore spike waveforms
superpositions.%
}:\begin{equation}
L\left(\mathcal{{D}},c\mid\theta\right)=\prod_{q=1}^{K}L_{l_{j}=q}\left(\mathcal{{D}},c\mid\theta\right).\label{eq:Multi-Neuron Likelihood}\end{equation}

\end{enumerate}

\subsubsection{Posterior density}

Now that we know how to write the likelihood of our \emph{augmented}
data set, we can obtain the posterior density of all our unknowns
$\Theta$ \emph{and} \emph{C} applying Bayes rule:\begin{equation}
\pi_{posterior}\left(\theta,c\,\mid\mathcal{{D}}\right)=\frac{L\left(\mathcal{{D}},c\mid\theta\right)\cdot\pi_{prior}\left(\theta\right)}{Z},\label{eq:Posterior PDF}\end{equation}

with the normalizing constant \emph{Z} given by:\begin{equation}
Z=\sum_{c\,\in\,\mathcal{{C}}}\int_{\theta}\, d\theta\, L\left(\mathcal{{D}},c\mid\theta\right)\cdot\pi_{prior}\left(\theta\right),\label{eq:Normalizing Constant}\end{equation}

where $\mathcal{{C}}$ is the set of all configurations.

\subsection{Remarks on the use of the posterior}

It should be clear that if we manage to get the posterior: $\pi_{posterior}\left(\theta,c\,\mid\mathcal{{D}}\right)$,
then we can answer any quantitative question about the data. For instance
if we want the probability of a specific configuration \emph{c} we
simply need to integrate out $\theta$:\begin{equation}
\pi\left(c\,\mid\mathcal{{D}}\right)=\int_{\theta}\, d\theta\,\pi_{posterior}\left(\theta,c\,\mid\mathcal{{D}}\right).\label{eq:Configuration Density}\end{equation}

If we are interested in the cross - correlogram between to neurons,
say 2 and 4, of the model and if we formally write $Cross_{2,4}\left(c\right)$
the function which takes as argument a specific configuration (\emph{c})
and returns the desired cross - correlogram, then the \emph{best}
estimate we can make of this quantity is:\begin{equation}
\left\langle Cross_{2,4}\right\rangle =\sum_{c\,\in\,\mathcal{{C}}}\int_{\theta}\, d\theta\,\pi_{posterior}\left(\theta,c\,\mid\mathcal{{D}}\right)\cdot Cross_{2,4}\left(c\right),\label{eq:Cross-Correlogram}\end{equation}

where we use the symbol $\left\langle \right\rangle $ to designate
averaging with respect to a proper \emph{pdf}.

\subsection{The normalizing constant problem and its solution}

\subsubsection{The problem}

If we take a look at our normalizing constant \emph{Z} (Eq. \ref{eq:Normalizing Constant})
we see a practical problem emerging. To compute \emph{Z} we need to
carry out a rather high dimensional integration (on $\theta$) \emph{and}
a summation over every possible configuration in $\mathcal{{C}}$
. That's where the serious problem arises because $\mathcal{{C}}$
contains $K^{N}$ elements! Blankly stated, we have no way, in real
situations where \emph{K} varies from 2 to 20 and where \emph{N} is
of the order of 1000, to compute \emph{Z}.

Fine, if we can't compute directly \emph{Z}, let us look for a way
around it. First, we introduce an {}``energy function'':\begin{equation}
E\left(\theta,c\,\mid\mathcal{{D}}\right)=-\ln\left[L\left(\mathcal{{D}},c\mid\theta\right)\cdot\pi_{prior}\left(\theta\right)\right].\label{eq:Energy Definition}\end{equation}

\emph{E} is defined in such a way that given the data ($\mathcal{{D}}$
), a value for the model parameters ($\theta$) and a configuration
(\emph{c}) we can compute it.

Second, we rewrite our posterior expression as follows:\begin{equation}
\pi_{posterior}\left(\theta,c\,\mid\mathcal{{D}}\right)=\frac{\exp\left[-\beta\, E\left(\theta,c\,\mid\mathcal{{D}}\right)\right]}{Z},\label{eq:Posterior PDF 2}\end{equation}

where $\beta=1$.

Third, we make an analogy with statistical physics: \emph{E} is an
energy, \emph{Z} a \emph{partition function}, $\beta$ is proportional
to the inverse of the temperature ($\left(kT\right)^{-1}$) and $\pi_{posterior}\left(\theta,c\,\mid\mathcal{{D}}\right)$
is the canonical distribution \cite{NewmanBarkema1999,LandauBinder2000}.

Fourth, we look at the way physicists cope with the combinatorial
explosion in the partition function...

\subsubsection{Its solution}

Physicists are interested in computing expected values (\emph{i.e.},
things that can be measured experimentally) like the expected internal
energy of a system:\begin{equation}
\left\langle E\right\rangle =\sum_{c\,\in\,\mathcal{{C}}}\int_{\theta}\, d\theta\,\frac{\exp\left[-\beta\, E\left(\theta,c\,\mid\mathcal{{D}}\right)\right]}{Z}\cdot E\left(\theta,c\,\mid\mathcal{{D}}\right),\label{eq:Energy-Expected-Value}\end{equation}

or like the pair - wise correlation function between two molecules
in a gas or in a fluid which will give rise to equations like Eq.
\ref{eq:Cross-Correlogram}. Of course they are very rarely able to
compute \emph{Z} explicitly so the trick they found fifty years ago
\cite{M(RT)2_1953} is to \emph{estimate} quantities like $\left\langle E\right\rangle $
with \emph{Monte Carlo} integration. 

At this point to make notation lighter we will call \emph{state} of
our {}``system'' the pair $\left(\theta,c\right)$ and we will write
it as: $x=\left(\theta,c\right)$. We will moreover drop the explicit
dependence on $\mathcal{{D}}$ in \emph{E}. That is, we will write
$E\left(x\right)$ for $E\left(\theta,c\,\mid\mathcal{{D}}\right)$.
Now by \emph{Monte Carlo} integration we mean that we {}``create''
a sequence of states: $x^{\left(1\right)},x^{\left(2\right)},\ldots,x^{\left(M\right)}$
such that:\begin{equation}
\lim_{M\rightarrow\infty}\overline{E}=\left\langle E\right\rangle ,\label{eq:Empirical To Expected Convergence}\end{equation}

where:\begin{equation}
\overline{E}=\frac{1}{M}\,\sum_{t=1}^{M}E\left(x^{\left(t\right)}\right),\label{eq:Energy Empirical Average}\end{equation}

is the empirical average of the energy. Showing how to create the
sequence of states $\left\{ x^{\left(t\right)}\right\} $ and proving
Eq. \ref{eq:Empirical To Expected Convergence} will keep us busy
in the next section.

\section{Markov Chains}

Before embarking with the details of Monte Carlo integration let's
remember the warning of one of its specialists, Alan Sokal \cite{Sokal1996}:

\begin{quotation}
Monte Carlo is an extremely bad method; it should be used only when
all alternative methods are worse.
\end{quotation}
That being said it seems we are in a situation with no alternative
method, so let's go for it.

Our first problem is that we can't directly (or independently) generate
the $x^{\left(t\right)}$ from the posterior density because this
density is too complicated. What we will do instead is generate the
sequence as a realization of a \emph{Markov Chain}. We will moreover
build the transition matrix of the Markov chain such that regardless
of the initial state $x^{\left(0\right)}$ , we will have for any
bounded function \emph{H} of $x$ the following limit:\begin{equation}
\lim_{M\rightarrow\infty}\frac{1}{M}\,\sum_{t=1}^{M}H\left(x^{\left(t\right)}\right)=\left\langle H\right\rangle .\label{eq:General Empirical to Expected Convergence}\end{equation}

In principle we should consider here Markov chains in a (partly) continuous
space because $x=\left(\theta,c\right)$ and $\theta$ is a continuous
vector. But we will take the easy solution, which is fundamentally
correct anyway, of saying that because we use a computer to simulate
the Markov chain, the best we can do is approximate the continuous
space in which $\theta$ lives by a \emph{discrete and finite} space.
We therefore split the study of the conditions under which Eq. \ref{eq:General Empirical to Expected Convergence}
is verified in the general state space (a product of a discrete and
a continuous space) into two questions%
\footnote{The results which will be presented in the next section have of course
equivalents in the general case. The reference which is always cited
by the serious guys is \cite{MeynTweedie1996}.%
}:

\begin{enumerate}
\item Under what conditions is Eq. \ref{eq:General Empirical to Expected Convergence}
verified for a Markov chain defined on a \emph{discrete and finite}
state space?
\item What kind of error do we produce by approximating our general state
space (a product of a discrete and of a continuous state space) by
a discrete one?
\end{enumerate}
We will answer here the first of these questions. For the second we
will hope (and check) that our discrete approximation is not too bad.

\subsection{Some notations and definitions}

Now that we are dealing with a discrete and finite state space we
label all our states $x$ with an integer and our state space $\mathcal{{X}}$
can be written as:\begin{equation}
\mathcal{{X}}=\left\{ x_{1},\ldots,x_{\nu}\right\} .\label{eq:State Space}\end{equation}

We can, moreover, make a unique correspondence between a random variable
(\emph{rv}), \emph{X}, and its \emph{probability mass function} $\pi_{X}$
that we can view as a row vector:\begin{equation}
\pi_{X}=\left(\pi_{X,1},\pi_{X,2},\ldots,\pi_{X,\nu}\right),\label{eq:PMF For RV}\end{equation}

where:\begin{equation}
\pi_{X,i}=Pr\left(X=x_{i}\right)\,,\,\forall x_{i}\in\mathcal{{X}}.\label{eq:PMF Components}\end{equation}

Strictly speaking a \emph{rv} would not be defined on a set like $\mathcal{{X}}$
but on a sample space $\mathcal{{S}}$ which would include every possible
outcome, that is: all elements of $\mathcal{{X}}$ as well as every
possible combination of them (like $x_{i}\cup x_{j}$, that we write
$x_{i}+x_{j}$ if $i\neq j$ because then $x_{i}$ and $x_{j}$ are
mutually exclusive) and the empty set $\textrm{\textrm{Ø}}$ (corresponding
to the impossible outcome). It is nevertheless clear that for a finite
state space, knowing the \emph{pmf} of a \emph{rv} is everything one
needs to compute the probability of any outcome (\emph{e.g.}, $Pr\left(x_{i}+x_{j}\right)=\pi_{X,i}+\pi_{X,j}$
). We will therefore in the sequel say that \emph{rv}s are defined
on the state space $\mathcal{{X}}$ acknowledging it is not the most
rigorous way to speak.

\paragraph{Formal definition of a Markov chain}

A sequence of random variables $X^{\left(0\right)},X^{\left(1\right)},\ldots$
, defined on a finite state space $\mathcal{{X}}$ is called a \emph{Markov
chain} if it satisfies the \emph{Markov property}:\begin{equation}
Pr\left(X^{\left(t+1\right)}=y\mid X^{\left(t\right)}=x,\ldots,X^{\left(0\right)}=z\right)=Pr\left(X^{\left(t+1\right)}=y\mid X^{\left(t\right)}=x\right).\label{eq:Markov Property}\end{equation}

If $Pr\left(X^{\left(t+1\right)}=y\mid X^{\left(t\right)}=x\right)$
does not explicitly depend on \emph{t} the chain is said to be \emph{homogeneous}.

An homogeneous Markov chain on a finite state space is clearly fully
described by its initial random variable $X^{\left(0\right)}$ and
by its \emph{transition matrix}%
\footnote{\textbf{Be careful here, the notations are, for historical reasons,
misleading. We write down the matrix: $T\left(x,y\right)$ where}
\textbf{\emph{x}} \textbf{is the state from which we start and} \textbf{\emph{y}}
\textbf{the state in which we end. When we use the probability transition
kernel notation $Pr\left(y\mid x\right)$ the starting state is on
the right side, the end site is on the left side! In this section
where we will discuss theoretical properties of Markov chains, we
will use the matrix notation which can be recognized by the the {}``,''
separating the start end end states. When later on, we will use the
kernel notation, the 2 states will be separated by the {}``$\mid$''
(and the end will be on the left side while the start will be on the
right).}%
} $T(x,y)=Pr\left(X^{\left(t+1\right)}=y\mid X^{\left(t\right)}=x\right)$.
This transition matrix must be \emph{stochastic} which means it must
satisfy (Sec \ref{sub:Stochastic Property of MC}):\begin{equation}
\sum_{y\,\in\,\mathcal{{X}}}T(x,y)=1,\:\forall\, x\,\in\,\mathcal{{X}}.\label{eq:Stochastic Matrix Property}\end{equation}

If \emph{m} is the \emph{pmf} of a \emph{rv} defined on $\mathcal{{X}}$,
we can easily show that the row vector $n$ defined as:\[
n_{j}=\sum_{i=1}^{\nu}m_{i}T(x_{i},x_{j}),\]

is itself a \emph{pmf} (\emph{i.}e\emph{.}, $0\leq n_{i}\leq1\,,\,\forall\, i\in\left\{ 1,\ldots,\nu\right\} $
and \emph{$\sum_{i=1}^{\nu}n_{i}=1$ )} and must therefore correspond
to a \emph{rv}. That justifies the following vector notation:\begin{equation}
\pi_{X^{\left(t+1\right)}}=\pi_{X^{\left(t\right)}}T,\label{eq:One Markov Step}\end{equation}

where $\pi_{X^{\left(t\right)}}$ and $\pi_{X^{\left(t+1\right)}}$
are the \emph{pmf} associated to two successive \emph{rv}s of a Markov
chain.

\paragraph{Stationary probability mass function}

A \emph{pmf} $m$ is said to be stationary with respect to the transition
matrix \emph{T} if: $mT=m$.

Applied to our Markov chain that means that if the \emph{pmf} of our
initial \emph{rv} ($\pi_{X^{\left(0\right)}}$ ) is stationary with
respect to \emph{T}, then the chain does not move (\emph{i.e.}, $X^{\left(t+1\right)}\equiv X^{\left(t\right)}\,,\, t\geq0$
).

The notion of stationary \emph{pmf} introduced we can restate our
convergence problem (Eq \ref{eq:General Empirical to Expected Convergence}).
What we want in practice is a Markov transition matrix, \emph{T},
which admits a single stationary \emph{pmf} (which would be in our
case $\pi_{posterior}$) and such that for any $\pi_{X^{\left(0\right)}}$
we have:\[
\lim_{t\rightarrow\infty}\pi_{X^{\left(t\right)}}=\pi_{posterior},\]

which means:\begin{equation}
\forall\, x\in\mathcal{{X}}\,,\,\forall\,\epsilon>0\,,\,\exists\, t\,\geq0\,:\,\left|\pi_{posterior}\left(x\right)-\pi_{X^{\left(t\right)}}\left(x\right)\right|\leq\epsilon.\label{eq:PMF Convergence Criterium}\end{equation}

It is clear that if we knew explicitly a candidate transition matrix,
\emph{T}, we could test that the posterior \emph{pmf} is stationary
and then using eigenvalue decomposition we could check if the largest
eigenvalue is 1, etc... See for instance: \cite{Bremaud1998,Fishman1996,RobertCasella1999,Liu2001,Neal1993,Sokal1996,NewmanBarkema1999}.
The problem is that we won't in general be able to write down explicitly
\emph{T}. We need therefore conditions we can check in practice and
which ensure the above convergence. These conditions are \emph{irreducibility}
and \emph{aperiodicity}.

\paragraph{Irreducible and aperiodic states}

(After Liu \cite{Liu2001}) A state $x\in\mathcal{{X}}$ is said to
be irreducible if under the transition rule one has nonzero probability
of moving from $x$ to any other state and then coming back in a finite
number of steps. A state $x\in\mathcal{{X}}$ is said to be aperiodic
if the greatest common divider of $\left\{ t\,:\, T^{t}\left(x,x\right)>0\right\} $
is 1.

It should not be to hard to see at this point that if one state $x\in\mathcal{{X}}$
is aperiodic, then all states in $\mathcal{{X}}$ must also be. Also,
if one state $y\in\mathcal{{X}}$ is aperiodic and the chain is irreducible,
then every state in $\mathcal{{X}}$ must be aperiodic.

\subsubsection*{Exercise 4}

Prove the validity of Eq. \ref{eq:Stochastic Matrix Property} for
a Markov chain on a finite state space $\mathcal{{X}}$ (Eq. \ref{eq:State Space}).
See Sec. \ref{sub:Stochastic Property of MC} for a solution.

\subsection{The fundamental theorem and the ergodic theorem}

\subsubsection*{Lemma}

Let $\left(X^{\left(0\right)},X^{\left(1\right)},\ldots\right)$ be
an irreducible and aperiodic Markov chain with state space $\mathcal{{X}}=\left\{ x_{1},x_{2},\ldots,x_{\nu}\right\} $
and transition matrix \emph{T}. Then there exists an $M<\infty$ such
that $T^{n}(x_{i},x_{j})>0$ for all $i,j\in\left\{ 1,\ldots,\nu\right\} $
and $n\geq M$.

\smallskip

\subsubsection*{Proof}

(After H\"aggstr\"om \cite{Haeggstroem2002}) By definition of irreducibility
and aperiodicity, there exist an integer $N<\infty$ such that $T^{n}\left(x_{i},x_{i}\right)>0$
for all $i\in\left\{ 1,\ldots,\nu\right\} $ and all $n\geq N$. Fix
two states $x_{i},x_{j}\in\mathcal{{X}}$ . By the assumed irreducibility,
we can find some $n_{i,j}$ such that $T^{n_{i,j}}\left(x_{i},x_{j}\right)>0$.
Let $M_{i,j}=N+n_{i,j}$. For any $m\geq M_{i,j}$, we have:\[
Pr\left(X^{\left(m\right)}=x_{j}\mid X^{\left(0\right)}=x_{i}\right)\geq Pr\left(X^{\left(m-n_{i,j}\right)}=x_{i},X^{\left(m\right)}=x_{j}\mid X^{\left(0\right)}=x_{i}\right),\]

and:\[
Pr\left(X^{\left(m-n_{i,j}\right)}=x_{i},X^{\left(m\right)}=x_{j}\mid X^{\left(0\right)}=x_{i}\right)=Pr\left(X^{\left(m-n_{i,j}\right)}=x_{i}\mid X^{\left(0\right)}=x_{i}\right)\, Pr\left(X^{\left(m\right)}=x_{j}\mid X^{\left(0\right)}=x_{i}\right)\]

but $m-n_{i,j}\geq0$ implies that: $Pr\left(X^{\left(m-n_{i,j}\right)}=x_{i}\mid X^{\left(0\right)}=x_{i}\right)>0$
and our choice of $n_{i,j}$ implies that: $Pr\left(X^{\left(m\right)}=x_{j}\mid X^{\left(0\right)}=x_{i}\right)>0$
therefore:\[
Pr\left(X^{\left(m\right)}=x_{j}\mid X^{\left(0\right)}=x_{i}\right)>0.\]

We have shown that $T^{m}\left(x_{i},x_{j}\right)>0$ for all $m\geq M_{i,j}$.
The lemma follows with:\[
M=\max\left\{ M_{1,1},M_{1,2},\ldots,M_{1,\nu},M_{2,1},\ldots,M_{\nu,\nu}\right\} .\]

\smallskip

\paragraph*{Remark}

It should be clear that the reciprocal of the lemma is true, that
is, if there exists an $M<0$ such that $T^{n}(x_{i},x_{j})>0$ for
all $i,j\in\left\{ 1,\ldots,\nu\right\} $ and $n\geq M$ then the
chain is irreducible and aperiodic. A transition matrix which has
all its elements strictly positive, or such that one of its powers
has all its elements strictly positive, will therefore give rise to
irreducible and aperiodic Markov chains.

We can now state and prove our first {}``fundamental'' theorem which
shows that irreducible and aperiodic Markov chains on finite state
spaces enjoy \emph{geometric convergence} to their \emph{unique} stationary
\emph{pmf}.

\subsubsection*{Fundamental theorem}

If an irreducible and aperiodic homogeneous Markov chain on a finite
state space $\mathcal{{X}}=\left\{ x_{1},x_{2},\ldots,x_{\nu}\right\} $
with a transition matrix $T\left(x,y\right)$ has $\pi_{stationary}$
as a stationary distribution then regardless of the initial \emph{pmf}
$\pi_{X^{\left(0\right)}}$ we have: \begin{equation}
\lim_{t\rightarrow\infty}\pi_{X^{\left(t\right)}}=\pi_{stationary}\label{eq:Convergence To Stationary}\end{equation}

and there exists an $1\geq\epsilon>0$ such that for all $x\in\mathcal{{X}}$
and all $t>1$ we have: \begin{equation}
\left|\pi_{stationary}\left(x\right)-\pi_{X^{\left(t\right)}}\left(x\right)\right|\leq\left(1-\epsilon\right)^{t}\label{eq:Geometric Convergence}\end{equation}

\smallskip

\subsubsection*{Proof}

Adapted from Neal \cite{Neal1993}. We can without loss of generality
assume that $T\left(x,y\right)>0\,,\,\forall\, x,y\,\in\,\mathcal{{X}}$.
If such is not the case for our initial transition matrix, we can
always use the preceding lemma to find an $M$ such that: $T'\left(x,y\right)=T^{M}\left(x,y\right)>0$
and work with the Markov chain whose transition matrix is $T'$ .
Then the uniqueness of the stationary \emph{pmf} regardless of the
initial \emph{pmf} will imply that the convergence of the {}``new''
Markov chain holds for the initial Markov chain as well.

So let's define:\[
\epsilon=\min_{x,y\,\in\,\mathcal{{X}}}\frac{T\left(x,y\right)}{\pi_{stationary}\left(y\right)}>0,\]

where we are assuming that $\pi_{stationary}\left(y\right)>0,$$\,\forall\, y\,\in\,\mathcal{{X}}$
(if not we just have to redefine the state space to get rid of $y$
and to take out row and column $y$ from $T$ ). Then from the transition
matrix definition and the stationarity assumption we have: $\pi_{stationary}\left(y\right)=\sum_{x}\pi_{stationary}\left(x\right)T\left(x,y\right),\,\forall\, y\,\in\,\mathcal{{X}}$
which with the above assumption gives:\[
1=\sum_{x}\pi_{stationary}\left(x\right)\frac{T\left(x,y\right)}{\pi_{stationary}\left(y\right)}\geq\epsilon\,\sum_{x}\pi_{stationary}\left(x\right),\]

it follows that: $\epsilon\leq1$. We will now show by induction that
the following equality holds:\begin{equation}
\pi_{X^{\left(t\right)}}\left(x\right)=\left[1-\left(1-\epsilon\right)^{t}\right]\pi_{stationary}\left(x\right)+\left(1-\epsilon\right)^{t}r_{t}\left(x\right)\:,\:\forall\, x\,\in\,\mathcal{{X}},\label{eq:Difference Between t and Stationary}\end{equation}

where $r_{t}$ is a \emph{pmf}.

This equality holds for $t=0$, with $r_{0}=\pi_{X^{\left(0\right)}}$.
Let us assume it is correct for $t$ and see what happens for $t+1$.
We have:\begin{eqnarray*}
\pi_{X^{\left(t+1\right)}}(y) & = & \sum_{x}\pi_{X^{\left(t\right)}}(x)\, T(x,y)\\
 & = & \left[1-\left(1-\epsilon\right)^{t}\right]\sum_{x}\pi_{stationary}(x)\, T(x,y)\:+\:\left(1-\epsilon\right)^{t}\sum_{x}r_{t}(x)\, T(x,y)\\
 & = & \left[1-\left(1-\epsilon\right)^{t}\right]\pi_{stationary}(y)\:+\:\left(1-\epsilon\right)^{t}\sum_{x}r_{t}(x)\,\left[T(x,y)-\epsilon\pi_{stationary}(y)+\epsilon\pi_{stationary}(y)\right]\\
 & = & \left[1-\left(1-\epsilon\right)^{t+1}\right]\pi_{stationary}(y)\:+\:\left(1-\epsilon\right)^{t}\sum_{x}r_{t}(x)\,\left[T(x,y)-\epsilon\pi_{stationary}(y)\right]\\
 & = & \left[1-\left(1-\epsilon\right)^{t+1}\right]\pi_{stationary}(y)\:+\:\left(1-\epsilon\right)^{t+1}\sum_{x}r_{t}(x)\,\frac{T(x,y)-\epsilon\pi_{stationary}(y)}{\left(1-\epsilon\right)}\\
 & = & \left[1-\left(1-\epsilon\right)^{t+1}\right]\pi_{stationary}(y)\:+\:\left(1-\epsilon\right)^{t+1}r_{t+1}\left(y\right)\end{eqnarray*}

where:\[
r_{t+1}\left(y\right)=\left[\sum_{x}r_{t}(x)\,\frac{T(x,y)}{\left(1-\epsilon\right)}\right]-\frac{\epsilon}{1-\epsilon}\,\pi_{stationary}\left(y\right)\]

One can easily check that: $r_{t+1}\left(y\right)\geq0,\,\forall\, y\,\in\,\mathcal{{X}}$
and that: $\sum_{y}r_{t+1}\left(y\right)=1$. That is, $r_{t+1}$
is a proper \emph{pmf} on $\mathcal{{X}}$.

With Eq \ref{eq:Difference Between t and Stationary} we can now show
that Eq \ref{eq:Convergence To Stationary} holds:\begin{eqnarray*}
\left|\pi_{stationary}(x)-\pi_{X^{\left(t\right)}}(x)\right| & = & \left|\pi_{stationary}(x)-\left[1-\left(1-\epsilon\right)^{t}\right]\pi_{stationary}(x)-\left(1-\epsilon\right)^{t}r_{t}\left(x\right)\right|\\
 & = & \left|\left(1-\epsilon\right)^{t}\pi_{stationary}(x)-\left(1-\epsilon\right)^{t}r_{t}\left(x\right)\right|\\
 & = & \left(1-\epsilon\right)^{t}\,\left|\pi_{stationary}(x)-r_{t}\left(x\right)\right|\\
 & \leq & \left(1-\epsilon\right)^{t}\end{eqnarray*}

\smallskip

\subsubsection*{Ergodic theorem}

Let $\left(X^{\left(0\right)},X^{\left(1\right)},\ldots\right)$ be
an irreducible and aperiodic homogeneous Markov chain on a finite
state space $\mathcal{{X}}=\left\{ x_{1},x_{2},\ldots,x_{\nu}\right\} $
with a transition matrix $T\left(x,y\right)$ and a stationary distribution
$\pi_{stationary}$. Let $a$ be a real valued function defined on
$\mathcal{{X}}$ and let $\overline{a}_{N}$ be the empirical average
of $a$ computed from a realization $\left(x^{\left(0\right)},x^{\left(1\right)},\ldots\right)$
of $\left(X^{\left(0\right)},X^{\left(1\right)},\ldots\right)$, that
is: $\overline{a}_{N}=\frac{1}{N}\,\sum_{t=0}^{N}a\left(x^{\left(t\right)}\right)$.
Then we have:\[
\lim_{N\rightarrow\infty}\left\langle \overline{a}_{N}\right\rangle =\left\langle a\right\rangle _{\pi_{stationary}},\]

where: \[
\left\langle a\right\rangle _{\pi_{stationary}}=\sum_{x\,\in\,\mathcal{{X}}}\pi_{stationary}\left(x\right)\, a\left(x\right),\]

and:\[
\left\langle \overline{a}_{N}\right\rangle =\frac{1}{N}\sum_{t=0}^{N}\,\sum_{x\,\in\,\mathcal{{X}}}\pi_{X^{\left(t\right)}}\left(x\right)\, a\left(x\right).\]

\smallskip

\subsubsection*{Proof}

Using Eq. \ref{eq:Difference Between t and Stationary} in the proof
of the fundamental theorem we can write:\begin{eqnarray*}
\left\langle \overline{a}_{N}\right\rangle  & = & \frac{1}{N}\sum_{t=0}^{N}\,\sum_{x\,\in\,\mathcal{{X}}}\pi_{X^{\left(t\right)}}\left(x\right)\, a\left(x\right)\\
 & = & \sum_{x\,\in\,\mathcal{{X}}}a\left(x\right)\,\frac{1}{N}\,\sum_{t=0}^{N}\pi_{X^{\left(t\right)}}\left(x\right)\\
 & = & \sum_{x\,\in\,\mathcal{{X}}}a\left(x\right)\,\left[\pi_{stationary}\left(x\right)\frac{1}{N}\sum_{t=0}^{N}\left(1-\zeta^{t}\right)\,+\,\frac{1}{N}\sum_{t=0}^{N}\zeta^{t}r_{t}\left(x\right)\right]\\
 & = & \left[\sum_{x\,\in\,\mathcal{{X}}}\pi_{stationary}\left(x\right)\, a\left(x\right)\right]\,+\,\left[\sum_{x\,\in\,\mathcal{{X}}}a\left(x\right)\,\left(\frac{1}{N}\sum_{t=0}^{N}\zeta^{t}\left(r_{t}\left(x\right)-\pi_{stationary}\left(x\right)\right)\right)\right]\end{eqnarray*}

where $\zeta$ corresponds to $1-\epsilon$ in Eq. \ref{eq:Difference Between t and Stationary}
and therefore: $0\leq\zeta<1$. To prove the ergodic theorem we just
need to show that the second term of the right hand side of the above
equation converges to 0 as $N\rightarrow\infty$. For that it is enough
to show that its modulus goes to 0:\begin{eqnarray*}
\left|\sum_{x\,\in\,\mathcal{{X}}}a\left(x\right)\,\left(\frac{1}{N}\sum_{t=0}^{N}\zeta^{t}\left(r_{t}\left(x\right)-\pi_{stationary}\left(x\right)\right)\right)\right| & \leq & \sum_{x\,\in\,\mathcal{{X}}}\left|a\left(x\right)\right|\,\left(\frac{1}{N}\sum_{t=0}^{N}\zeta^{t}\left|r_{t}\left(x\right)-\pi_{stationary}\left(x\right)\right|\right)\\
 & \leq & \sum_{x\,\in\,\mathcal{{X}}}\left|a\left(x\right)\right|\,\frac{1}{N}\sum_{t=0}^{N}\zeta^{t}\end{eqnarray*}
 We just need to recognize here the geometrical series $\sum_{t=0}^{N}\zeta^{t}$
which converges to $\frac{1}{1-\zeta}$ when $N\rightarrow\infty$.
That completes the proof.

\smallskip

\paragraph{What did we learn in this section?}

For a given \emph{pmf $\pi$} defined on a finite state space $\mathcal{{X}}$
if we can find an irreducible aperiodic Markov matrix \emph{T} which
admits $\pi$ as its (necessarily only) stationary distribution, then
\emph{regardless} of our initial choice of \emph{pmf} $\pi_{0}$ defined
on $\mathcal{{X}}$ and regardless of the function $a:\,\mathcal{{X}}\rightarrow\Re$
, an \emph{asymptotically unbiased} estimate of $\left\langle a\right\rangle =\sum_{x\,\in\,\mathcal{{X}}}a\left(x\right)\,\pi\left(x\right)$
can be obtained by simulating the Markov chain to get a sequence of
states: $\left(x^{\left(0\right)},x^{\left(1\right)},\ldots,x^{\left(N\right)}\right)$
and computing the empirical average: $\overline{a}=\frac{1}{N}\sum_{t=0}^{N}a\left(x^{\left(t\right)}\right)$.

\section{The Metropolis - Hastings algorithm and its relatives}

What we need now is a way to construct the transition matrix \emph{T}
and more specifically a method which works with an \emph{unnormalized}
version of the stationary density. We first introduce the notion of
\emph{detailed balance}.

\paragraph*{Detailed balance definition\label{par:Detailed-balance-definition}}

We say that a \emph{pmf} $\pi$ defined on a (finite) state space
$\mathcal{{X}}$ and a Markov transition matrix \emph{T} satisfy the
\emph{detailed balance} if: $\forall x,y\,\in\,\mathcal{{X}},\:\pi\left(x\right)T\left(x,y\right)=\pi\left(y\right)T\left(y,x\right)$.

\paragraph*{Detailed balance theorem\label{par:Detailed-balance-theorem}}

If the \emph{pmf} $\pi$ defined on a (finite) state space $\mathcal{{X}}$
and a Markov transition matrix \emph{T} satisfy the \emph{detailed
balance} then $\pi$ is stationary for \emph{T}.

\subsubsection*{Exercise 5}

Prove the theorem. See Sec. \ref{sub:Detailed Balance} for solution.

\subsection{The Metropolis - Hastings algorithm}

\subsubsection{Second fundamental theorem\label{sub:Second-fundamental-theorem}}

Let $\pi$ be a \emph{pmf} defined on a (finite) state space $\mathcal{{X}}$
and \emph{T} and \emph{G} two Markov transition matrices defined on
$\mathcal{{X}}$ satisfying:\begin{eqnarray*}
T\left(x,y\right) & = & A\left(x,y\right)G\left(x,y\right)\; if\, x\neq y\\
T\left(x,x\right) & = & 1-\sum_{y\,\in\,\mathcal{{X}},y\neq x}T(x,y)\end{eqnarray*}

where $G\left(y,x\right)=0$ if $G\left(x,y\right)=0$ and\begin{eqnarray}
A\left(x,y\right) & = & \min\left(1,\frac{\pi\left(y\right)G\left(y,x\right)}{\pi\left(x\right)G\left(x,y\right)}\right),\; if\, G\left(x,y\right)>0\label{eq:Acceptance Probability}\\
 & = & 1,\quad otherwise\nonumber \end{eqnarray}

then

\begin{itemize}
\item $\pi$ and \emph{T} satisfy the detailed balance condition
\item $\pi T=\pi$
\item if \emph{G} is irreducible and aperiodic so is \emph{T}
\end{itemize}

\subsubsection*{Exercise 6}

Prove the theorem.

\smallskip

\paragraph{THE MOST IMPORTANT COMMENT OF THESE LECTURES NOTES}

\begin{quote}
This theorem is exactly what we were looking for. It tells us how
to modify an irreducible and aperiodic Markov transition (\emph{G})
\emph{}such that a \emph{pmf} $\pi$ of our choice will be the stationary
distribution and it does that by requiring a knowledge of the desired
stationary $\pi$ only up to a normalizing constant, because Eq. \ref{eq:Acceptance Probability}
involves the ratio of two values of $\pi$. 
\end{quote}
\smallskip

\emph{G} is often called the \emph{proposal} transition and \emph{A}
the \emph{acceptance} probability. The \emph{Metropolis algorithm}
\cite{M(RT)2_1953} (sometimes called the $M\left(RT\right)^{2}$
algorithm because of its authors names) is obtained with a symmetrical
\emph{G} (\emph{i.e.}, $G\left(x,y\right)=G\left(y,x\right)$). The
above formulation (in fact a more general one) is due to Hastings
\cite{Hastings1970}. The set of techniques which involves the construction
of a Markov chain (with the Metropolis - Hastings algorithm) to perform
Monte Carlo integration is called \emph{Dynamic Monte Carlo} by physicists
\cite{Sokal1996,NewmanBarkema1999,LandauBinder2000} and \emph{Markov
Chain Monte Carlo} (\emph{MCMC}) by statisticians \cite{Geman1984,RobertCasella1999,Liu2001}.

\subsection{Metropolis - Hastings and Gibbs algorithms for multi - dimensional
spaces\label{sub:Metropolis---Hastings-details}}

Talking of multi - dimensional spaces when we started by saying we
were working with a discrete and finite one can seem a bit strange.
It is nevertheless useful not to say necessary to keep a trace of
the multi - dimensionality of our {}``initial'' space (ideally all
our model parameters: $\textbf{{P}}_{k},\Delta_{k},\Lambda_{k},S_{k},F_{k}$
{}``live'' in continuous spaces). If not it would be extremely hard
to find our way on the map between the discrete approximations of
all these continuous spaces and the genuine discrete and finite space
on which our simulated Markov chain evolves.

We consider therefore now that our random variables are {}``vectors''
and $X^{\left(t\right)}$ becomes $\textbf{{X}}^{\left(t\right)}$.
We can think of it as follows: $\textbf{{X}}_{1}^{\left(t\right)}$
corresponds to $P_{1,1}^{\left(t\right)}$ the maximal peak amplitude
of the first neuron in the model on the first recording site after
\emph{t} steps,..., $\textbf{{X}}_{4}^{\left(t\right)}$corresponds
to $P_{1,4}^{\left(t\right)}$ the maximal peak amplitude of the first
neuron in the model on the fourth recording site (assuming we are
using tetrodes), $\textbf{{X}}_{5}^{\left(t\right)}$ corresponds
to the parameter $\Delta^{\left(t\right)}$ of the first neuron, $\textbf{{X}}_{6}^{\left(t\right)}$
corresponds to the parameter $\Lambda^{\left(t\right)}$ of the first
neuron, $\textbf{{X}}_{7}^{\left(t\right)}$ corresponds to the parameter
$S^{\left(t\right)}$ of the first neuron, $\textbf{{X}}_{8}^{\left(t\right)}$
corresponds the parameter $F^{\left(t\right)}$ of the first neuron,
$\textbf{{X}}_{9}^{\left(t\right)}$ corresponds to $P_{2,1}^{\left(t\right)}$
the maximal peak amplitude of the second neuron in the model on the
first recording site, etc. The problem is that it turns out to be
very hard to find a transition matrix \emph{G} acting on object like
these {}``new'' \emph{pmf}s: $\pi_{\textbf{{X}}^{\left(t\right)}}$such
that the acceptance probabilities (Eq. \ref{eq:Acceptance Probability})
are not negligible. The way around this difficulty is to build the
transition matrix \emph{T} as a combination of component - wise transition
matrices like $T_{1}$ which acts only on the first component of $\pi_{\textbf{{X}}^{\left(t\right)}}$,
$T_{2}$ which acts only on the second, etc. We just need to make
sure that our combination is irreducible, and aperiodic (we assume
here that we build each individual $T_{j}$ such that $\pi_{posterior}$
is its stationary \emph{pmf}). A way to do that is to construct each
$T_{j}$ such that it is irreducible and aperiodic on its {}``own''
sub - space which is obtained practically by building the matrix such
that it has a strictly positive probability to produce a move anywhere
on its sub - space. Then two combinations of these $T_{j}$s are typically
used, the first one being:\[
T=w_{1}T_{1}+w_{2}T_{2}+\ldots+w_{m}T_{m},\]

where \emph{m} is the number of components of the random vectors and
were the $w_{j}$s are components of a \emph{pmf} defined on the the
set of coordinates $\left\{ 1,\ldots,m\right\} $. It should be clear
that if each $T_{j}$ is irreducible and aperiodic on its own sub
- space, then $T$ will be irreducible and aperiodic on the {}``complete''
state space. Because each $T_{j}$ is built such that $\pi_{posterior}$
is its stationary \emph{pmf}, $\pi_{posterior}$ will be the stationary
\emph{pmf} of $T$ as well. The concrete implementation of this scheme
would go as follows: at each {}``sub - step'' a coordinate \emph{j}
is randomly drawn from \emph{w}, then a random move is performed using
$T_{j}$. It is customary to call {}``Monte Carlo step'' a sequence
of \emph{m} successive such {}``sub - steps'' (that means that on
average each model parameter will be {}``changed'' during a Monte
Carlo step).

The second combination is:\begin{equation}
T=T_{1}\times T_{2}\times\ldots\times T_{m},\label{eq:sequential-scheme}\end{equation}

that is, each model parameter is successively {}``changed''. In
the same way as above the irreducibility and aperiodicity of the $T_{j}$s
in their sub - spaces will give rise to an irreducible and aperiodic
$T$ on the {}``full'' parameter space. The main difference is that
detailed balance condition which be imposed by construction to the
pairs $T_{j}$, $\pi_{posterior}$ is not kept by the pair $T$, $\pi_{posterior}$.
We only have the stationarity property (which is enough to ensure
convergence). Of course variations on that scheme can be used like
using random permutations of the $T_{j}$s (which would restore the
detailed balance condition for the pair $T$, $\pi_{posterior}$).
A {}``Monte Carlo step'' for those schemes is obtained after the
application of the complete series of $T_{j}$s. See \cite{Fishman1996}
for details.

\subsubsection{An example: the Gibbs sampler for the parameters of the \emph{ISI}
density}

The previous discussion seemed probably a bit abstract for most of
the readers. In order to be more precise about what we meant we will
start by considering the following {}``simple'' situation. Let's
assume that we are given a sample of 25 \emph{isi}s: $\mathcal{{D}}=\left\{ i_{1},\ldots,i_{25}\right\} $
drawn independently from a log-Normal density with parameters: $s_{actual},f_{actual}$.
We are asked for a Bayesian estimation of values of $s$ and $f$
(assuming flat priors for these parameters like in Exercise 3):\[
\pi_{isi,posterior}\left(s,f\mid\mathcal{{D}}\right)\,\alpha\, L_{isi}\left(\mathcal{{D}}\mid s,f\right)\cdot\pi_{isi,prior}\left(s,f\right),\]

where \emph{$L_{isi}$} is easily obtained from Eq. \ref{eq:Single Neuron Log-Likelihood ISI}:\begin{equation}
L_{isi}\left(\mathcal{{D}}\mid s,f\right)=\frac{1}{f^{N}}\exp\left[-\frac{1}{2}\,\frac{1}{f^{2}}\,\sum_{j=1}^{N}\left(\ln i_{j}-\ln s\right)^{2}\right]\label{eq:Single Neuron ISI Likelihood}\end{equation}
We do not recognize any classical \emph{pdf} in Eq. \ref{eq:Single Neuron ISI Likelihood}
and we choose to use a MCMC approach. Following the previous discussion
we will try to build our transition matrix \emph{T} as a {}``product''
$T_{S}\times T_{F}$. Where $T_{S}$ does only change \emph{s} and
$T_{F}$ does only change \emph{f} and where both are irreducible
and aperiodic on their own sub-space. According to the second fundamental
theorem we first need to find proposal transitions: $G_{s}\left(s_{now},s_{proposed}\mid f,\mathcal{{D}}\right)$
and $G_{f}\left(f_{now},f_{proposed}\mid s,\mathcal{{D}}\right)$.
But questions 1 and 3 of Exercise 3 provide us with such proposal
transitions. In fact these transitions are a bit special because they
do not depend on the present value of the parameter we will try to
change and because they indeed correspond to the posterior conditional
density of this parameter. A direct consequence of the latter fact
is that the acceptance probability (Eq. \ref{eq:Acceptance Probability})
is 1. An algorithm where the proposal transition for a parameter is
the posterior conditional of this parameter is called a \emph{Gibbs
sampler} by statisticians and a \emph{heat bath algorithm} by physicists.

We therefore end up with the following algorithm (using the results
of Exercise 3):

\begin{enumerate}
\item Chose randomly $s^{\left(0\right)}\in\left[s_{min},s_{max}\right]$
and $f^{\left(0\right)}\in\left[f_{min},f_{max}\right]$.
\item Given $f^{\left(t\right)}$ draw:\[
s^{\left(t+1\right)}\,\sim\,\pi\left(\quad\mid f^{\left(t\right)},\mathcal{{D}}\right)\]
\\
where $\pi\left(\quad\mid f^{\left(t\right)},\mathcal{{D}}\right)$
is defined by Eq. \ref{eq:Posterior Scale PDF} (remark that $s^{\left(t+1\right)}$
is independent of $s^{\left(t\right)}$).
\item Given $s^{\left(t+1\right)}$ draw:\[
f^{\left(t+1\right)}\,\sim\,\pi\left(\quad\mid s^{\left(t+1\right)},\mathcal{{D}}\right)\]
\\
where $\pi\left(\quad\mid s^{\left(t+1\right)},\mathcal{{D}}\right)$
is defined by Eq. \ref{eq:Posterior Shape PDF}.
\end{enumerate}

\subsubsection*{Exercise 7}

Simulate 25 \emph{isi} following a log-Normal density with values
of your choice for the pair $s_{actual},f_{actual}$. Implement the
algorithm and compare its output with the maximum likelihood based
inference. That is with the maximal likelihood estimates for \emph{s}
and \emph{f} (given by Eq. \ref{eq:Best Scale} \& \ref{eq:Best Shape}).
You should compute as well the Hessian of the log-likelihood function
at its maximum to get confidence intervals (see the chapter of E.
Brown).

\subsubsection{Generation of the amplitude parameters of our model\label{sub:Generation-of-the-Amplitude-Para}}

\begin{figure}
\begin{center}

\includegraphics[%
  scale=0.4,
  angle=270]{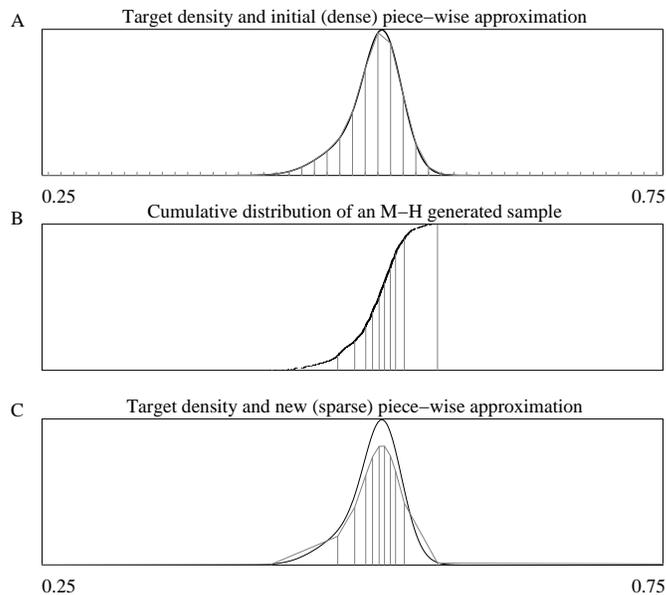}

\end{center}

\caption{\label{cap:Fig-Piece-Wise}A, a simple one-dimensional target density
(black curve, a Gaussian mixture: $0.3\,\mathcal{{N}}\left(0.5,0.025^{2}\right)+0.7\,\mathcal{{N}}\left(0.525,0.015^{2}\right)$
) together with its linear piece-wise approximation (gray curve) based
on 100 discrete samples (the prior density is supposed to be flat
between 0 and 1 and zero everywhere else). B, A sample of size 1000
is drawn using the piece-wise linear approximation as a proposal and
a MH acceptance rule. The cumulative distribution of the sample is
shown (black) together with the location of the 10 quantiles. C, Using
the location of the 10 quantiles, the boundaries of the prior and
the location of the smallest generated value, a sparser piece-wise
linear approximation of the target density is built.}
\end{figure}

By {}``amplitude parameters'' we mean here the following parameters:
$\textbf{{P}},\Delta,\Lambda$. Given a data set from a single neuron:
$\mathcal{{D}}=\left\{ \left(i_{1},\textbf{{a}}_{1}\right),\ldots,\left(i_{N},\textbf{{a}}_{N}\right)\right\} $
(see Sec. \ref{sub:Single-Neuron-Likelihood-Comp}) we now try to
perform Bayesian inference on all its parameters: $\textbf{{P}},\Delta,\Lambda,S,F$.
We again split our transition matrix \emph{T} into parameter specific
transitions: $T=T_{P_{1}}\times T_{P_{2}}\times T_{P_{3}}\times T_{P_{4}}\times T_{\Delta}\times T_{\Lambda}\times T_{S}\times T_{F}$.
We have seen in the previous example how to get $T_{S}$ and $T_{F}$.
Following the same line we could try to build a Gibbs sampler for
the other parameters as well. The problem is that the part of the
Likelihood function which depends on the amplitude parameters (obtained
from Eq. \ref{eq:Single Neuron Log-Likelihood Amplitude}):\[
L_{amp}\left(\mathcal{{D}}\mid\textbf{{p}},\delta,\lambda\right)=\exp\left[-\frac{1}{2}\sum_{j=1}^{N}\left\Vert \textbf{{a}}_{j}-\textbf{{p}}\cdot\left(1-\delta\cdot\exp\left(-\lambda\cdot i_{j}\right)\right)\right\Vert ^{2}\right]\]

does not correspond to any know \emph{pdf} even when considered as
a function of a single parameter, say $\delta$. The reader can notice
that such would not be the case if we had $\delta=0$ and if we knew
it, see \cite{NugyenEtAl2003}. A robust solution to this problem
is to use a piece-wise linear approximation of the posterior conditional
as a proposal transition for each parameter (\emph{e.g.}, $G_{\Delta}\left(\quad\mid\textbf{{p}},\lambda,s,f,\mathcal{{D}}\right)$)
and then an acceptance probability as defined by Eq. \ref{eq:Acceptance Probability}.
More specifically we can start with 101 regularly spaced {}``sampling
values'' of $\delta$:\[
\delta\in\left\{ \delta_{0}=0,\delta_{1}=0.01,\ldots,\delta_{99}=0.99,\delta_{100}=1\right\} ,\]

compute 101 values of:\begin{equation}
L_{amp}\left(\mathcal{{D}}\mid\textbf{{p}},\delta_{i},\lambda\right)\label{eq:Likelihood-of-Delta}\end{equation}

and define:\begin{eqnarray*}
G_{\Delta}\left(\delta\mid\textbf{{p}},\lambda,s,f,\mathcal{{D}}\right) & = & \mathcal{{N}}_{\Delta}\cdot\left[L_{amp}\left(\mathcal{{D}}\mid\textbf{{p}},\delta_{i},\lambda\right)+\frac{L_{amp}\left(\mathcal{{D}}\mid\textbf{{p}},\delta_{i+1},\lambda\right)-L_{amp}\left(\mathcal{{D}}\mid\textbf{{p}},\delta_{i},\lambda\right)}{\delta_{i+1}-\delta_{i}}\left(\delta-\delta_{i}\right)\right]\\
\end{eqnarray*}

where $\mathcal{{N}}_{\Delta}$ ensures that $G_{\Delta}$ is properly
normalized and where $\delta\in\left[\delta_{i},\delta_{i+1}\right]$.
The obvious problem with this approach is that we need to have a reasonably
good piece-wise linear approximation of the corresponding posterior
conditional \emph{pdf} in order to get reasonable values for our acceptance
probability. That means that when we start our Markov chain and we
do not have a precise idea of the actual shape of this posterior conditional
density we have to use a lot of sampling points. We spend therefore
a lot of time computing terms like Eq. \ref{eq:Likelihood-of-Delta}.
A very significant increase of the algorithm speed can thus be obtained
by using a first run with a lot of sampling points (say 101), then
reposition fewer sampling point (say 13)%
\footnote{We typically use 13 points because we want to have the 10 quantiles,
the 2 extrema (here, $\delta_{min}$ and $\delta_{max}$) allowed
by our priors and an extra point corresponding to the smallest value
obtained in our sample. The reader can {}``show'' as an exercise
the usefulness of this extra sample point. The easiest way to see
the use of this point is to try without it and to remember that the
piece-wise linear approximation \emph{must be normalized}.%
} using the output of the first run. That's explained in details in
\cite{PouzatEtAl2004} and illustrated on Fig. \ref{cap:Fig-Piece-Wise}.
When dealing with multiple neurons data, Eq. \ref{eq:Multi-Neuron Likelihood}
shows that the same approach can be immediately applied after the
introduction of the configuration.

\subsubsection{Generation of the configuration}

This can be done (almost) exactly as for a {}``normal'' Potts model
\cite{NewmanBarkema1999,Sokal1996} and is left as an exercise to
the reader (the answer can be found in \cite{PouzatEtAl2004}).

\subsubsection{Our complete MC step}

In Sec. \ref{sec:An-Example} we will give an example of our algorithm
at work with the following sequential MC step:\begin{equation}
T_{l_{1}}\times\ldots\times T_{l_{N}}\times T_{P_{1,1}}\times\ldots\times T_{f_{1}}\times\ldots\times T_{P_{K,1}}\times\ldots\times T_{f_{K}}\label{eq:Complete-MC-Step}\end{equation}

\section{Priors choice\label{sec:Choice-of-the-priors}}

We will assume here that we know {}``little'' \emph{a priori} about
our parameters values and that the joint prior density $\pi_{prior}(\theta)$
can be written as a product of the densities for each component of
$\theta$, that is:\[
\pi_{prior}(\theta)=\prod_{q=1}^{K}\pi(f_{q})\cdot\pi(s_{q})\cdot\pi(\delta_{q})\cdot\pi(\lambda_{q})\cdot\pi(P_{q,1})\cdot\pi(P_{q,2})\cdot\pi(P_{q,3})\cdot\pi(P_{q,4})\]

where we are assuming that four recording sites have been used. We
will further assume that our signal to noise ratio is not better 20
(a rather optimistic value), that our spikes are positive and therefore
the $\pi(P_{q,1\ldots4})$ are null below 0 and above +20 (remember
we are working with normalized amplitudes). We will reflect our absence
of prior knowledge about the amplitudes by taking a uniform distribution
between 0 and +20. The $\lambda$ value reported by Fee et al \cite{FeeEtAl1996}
is 45.5 $s^{-1}$. $\lambda$ must, moreover be smaller than $\infty$,
we adopt a prior density uniform between 10 and 200 $s^{-1}$. $\delta$
must be $\leq1$ (the amplitude of a spike from a given neuron on
a given recording site \emph{does not} change sign) and $\geq0$ (spikes
do not become larger upon short ISI), we therefore use a uniform density
between 0 and 1 for $\delta$. An inspection of the effect of the
shape parameter $f$ on the ISI density is enough to convince an experienced
neurophysiologist that empirical unimodal ISI densities from well
isolated neurons will have $f\in\left[0.1,2\right]$. We therefore
take a prior density uniform between 0.1 and 2 for $f$. The same
considerations leads us to take a uniform prior density between 0.005
and 0.5 for \emph{s}.

\section{The good use of the ergodic theorem. A warning.}

The ergodic theorem is the key theoretical result justifying the use
of Monte Carlo integration to solve tough problems. When using these
methods we should nevertheless be aware that the theorem applies only
when the number of Monte Carlo steps of our algorithms go to infinity%
\footnote{This is not even true because our algorithm use \emph{pseudo - random
- number} generators which among many shortcomings have a \emph{finite}
period. See Chap 7 of \cite{Fishman1996}, Chap 2 of \cite{RobertCasella1999}
and \cite{LEcuyer1994}.%
} and because such is never practically the case we will commit errors.
We would commit errors even if our draws where directly generated
from $\pi_{post}$. The difference between the MCMC/Dynamic MC based
estimates and estimates based on direct samples ({}``plain MC'')
is that the variance of the estimators of the former have to be corrected
to take into account the correlation of the states generated by the
Markov chain. We explain now how to do that in practice, for a theoretical
justification see Sokal \cite{Sokal1996} and Janke \cite{Janke2002}.

\subsection{Autocorrelation functions and confidence intervals\label{sub:Autocorrelation-functions-and}}

We have to compute for each parameter, $\theta_{i}$, of the model
the normalized autocorrelation function (ACF), $\rho_{norm}(l;\theta_{i})$,
defined by:\begin{eqnarray}
\rho(l;\theta_{i}) & = & \frac{1}{N_{T}-N_{D}-l}\cdot\sum_{t=N_{D}}^{N_{T}-l}(\theta_{i}^{(t)}-\bar{\theta}_{i})\cdot(\theta_{i}^{(t+l)}-\bar{\theta}_{i})\nonumber \\
\rho_{norm}(l;\theta_{i}) & = & \frac{\rho(l;\theta_{i})}{\rho(0;\theta_{i})}\label{eq:Autocorrelation}\end{eqnarray}

Where $N_{T}$ is the total number of MC steps performed and $N_{D}$
is the number of steps required to reach {}``equilibrium'' (see
Sec. \ref{sub:Initialization-bias}, \ref{sub:Algorithm-dynamics-NOREM}
\& \ref{sub:Algorithm-dynamics-REM}). Then we compute the \emph{integrated
autocorrelation time}, $\tau_{autoco}(\theta_{i})$:\begin{equation}
\tau_{autoco}(\theta_{i})=\frac{1}{2}+\sum_{l=1}^{L}\rho(l;\theta_{i})\label{eq:Autoco-time}\end{equation}
 where \emph{L} is the lag at which $\rho$ starts oscillating around
0. Using an empirical variance, $\sigma^{2}(\theta_{i})$ of parameter
$\theta_{i}$, defined in the usual way:\begin{equation}
\sigma^{2}(\theta_{i})=\frac{1}{N_{T}-N_{D}-1}\,\sum_{t=N_{D}}^{N_{T}}(\theta_{i}^{(t)}-\overline{\theta}_{i})^{2}\label{eq:Empirical-variance}\end{equation}
 where $\overline{\theta}_{i}$ is defined like in Eq. \ref{eq:Energy Empirical Average}.
Our estimate of the variance, $Var\left[\overline{\theta}_{i}\right]$
of $\overline{\theta}_{i}$ becomes:\begin{equation}
Var\left[\overline{\theta}_{i}\right]=\frac{2\,\tau_{autoco}(\theta_{i})}{N_{T}-N_{D}-1}\cdot\sigma^{2}(\theta_{i})\label{eq:Variance-estimate}\end{equation}
In the sequel, the confidence intervals on our parameters estimates
are given by the square root of the above defined variance (Table
\ref{cap:Para-Estimates-NOREM} \& \ref{cap:Para-Estimates-REM}). 

We can view the effect of the autocorrelation of the states of the
chain as a reduction of our effective sample size by a factor: $2\,\tau_{autoco}(\theta_{i})$.
This gives us a first quantitative element on which different algorithms
can be compared (remember that the MH algorithm gives us a lot of
freedom on the choice of proposal transition kernels). It is clear
that the faster the autocorrelation functions of the parameters fall
to zero, the greater the statistical efficiency of the algorithm.
The other quantitative element we want to consider is the computational
time, $\tau_{cpu}$, required to perform one MC step of the algorithm.
One could for instance imagine that a new sophisticated proposal transition
kernel allows us to reduce the largest $\tau_{autoco}$ of our standard
algorithm by a factor of 10, but at the expense of an increase of
$\tau_{cpu}$ by a factor of 100. Globally the new algorithm would
be 10 times less efficient than the original one. What we want to
keep as small as possible is therefore the product: $\tau_{autoco}\cdot\tau_{cpu}$.

\subsection{Initialization bias\label{sub:Initialization-bias}}

The second source of error in our (finite size) MC estimates is a
bias induced by the state $(\theta^{(0)},C^{(0)})$ with which the
chain is initialized \cite{Sokal1996,Fishman1996}. The bad news concerning
this source of error is that there is no general theoretical result
providing guidance on the way to handle it, but the booming activity
in the Markov chain field already produced encouraging results in
particular cases \cite{Liu2001}. The common wisdom in the field is
to monitor parameters (and labels) evolution, and/or functions of
them like the energy (Eq. \ref{eq:Energy Definition}). Based on examination
of evolution plots (\emph{eg}, Fig. \ref{cap:NRJ-EVOL-NOREM} \& \ref{cap:PEAK-EVOL-NOREM})
and/or on application of time-series analysis tools, the user will
decide that {}``equilibrium'' has been reached and discard the parameters
values before equilibrium. More sophisticated tests do exist \cite{RobertCasella1999}
but they wont be used in this chapter. These first two sources of
error, finite sample size and initialization bias, are clearly common
to all MCMC approaches.

\section{Slow relaxation and the Replica Exchange Method\label{sec:Slow-relaxation-and-REM}}

A third source of error appears only when the energy function exhibits
several local minima. In the latter case, the Markov chain realization
can get trapped in a local minimum which could be a poor representation
of the whole energy function. This sensitivity to local minima arises
from the local nature of the transitions generated by the MH algorithm.
That is, if we use a sequential scheme like Eq. \ref{eq:Complete-MC-Step},
at each MC time step, we first attempt to change the label of spike
1, then the one of spike 2, ..., then the one of spike N, then we
try to change the first component of $\theta$ ($P_{1,1}$), and so
on until the last component ($\sigma_{K}$). That implies that if
we start in a local minimum and if we need to change, say, the labels
of 10 spikes to reach another lower local minimum, we could have a
situation in which the first 3 label changes are energetically unfavorable
(giving, for instance, an acceptance probability, Eq. \ref{eq:Acceptance Probability},
of $0.1$ per label change) which would make the probability to accept
the succession of changes very low ( $0.1^{3}$ )... meaning that
our Markov chain would spend a long time in the initial local minimum
before {}``escaping'' to the neighboring one. Stated more quantitatively,
the average time the chain will take to escape from a local minimum
with energy $E_{min}$ grows as the exponential of the energy difference
between the energy, $E^{*}$, of the highest energy state the chain
has to go through to escape and $E_{min}$: \[
\tau_{escape}\,\alpha\,\exp\left[\beta\left(E^{*}-E_{min}\right)\right]\]

Our chains will therefore exhibit an Arrhenius behavior. To sample
more efficiently such state spaces, the Replica Exchange Method (REM)
\cite{HukushimaNemoto1996,Mitsutake2001}, also known as the Parallel
Tempering Method \cite{Hansman1997,YandePablo1999,FalcioniDeem1999},
considers \emph{R} replicas of the system with an increasing sequence
of temperatures (or a decreasing sequence of $\beta$) and a dynamic
defined by two types of transitions : usual MH transitions performed
independently on each replica according to the rule defined by Eq.
\ref{eq:Complete-MC-Step} and a replica exchange transition. The
key idea is that the high temperature (low $\beta$) replicas will
be able to easily cross the energy barriers separating local minima
(in the example above, if we had a probability of $0.1^{3}$ to accept
a sequence of labels switch for the replica at $\beta=1$, the replica
at $\beta=0.2$ will have a probability $0.1^{3\cdot0.2}\approx0.25$
to accept the same sequence). What is needed is a way to generate
replica exchange transitions such that the replica at $\beta$ = 1
generates a sample from $\pi_{post}$ defined by Eq. \ref{eq:Posterior PDF 2}.
Formally the REM consists in simulating, on an {}``extended ensemble'',
a Markov chain whose unique stationary density is given by:\begin{equation}
\pi_{ee}\left(\theta_{1},C_{1},\ldots,\theta_{R},C_{R}\right)=\pi_{post,\beta_{1}}\left(\theta_{1},C_{1}\right)\ldots\pi_{post,\beta_{R}}\left(\theta_{R},C_{R}\right)\label{eq:Joint-replica-density}\end{equation}

where {}``\emph{ee}'' in $\pi_{ee}$ stands for {}``extended ensemble''
\cite{Iba2001}, \emph{R} is the number of simulated replicas, $\beta_{1}>\ldots>\beta_{R}$
for convenience and:\begin{equation}
\pi_{post,\beta_{i}}\left(\theta_{i},C_{i}\right)=\frac{\exp\left[-\beta_{i}E\left(\theta_{i},C_{i}\right)\right]}{Z\left(\beta_{i}\right)}\label{eq:Posterior-beta-dependent}\end{equation}

That is, compared to Eq. \ref{eq:Posterior PDF 2}, we now explicitly
allow $\beta$ to be different from 1. To construct our {}``complete''
transition kernel we apply our previous procedure. That is, we construct
it as a sequence of parameter, label and inter-replica specific MH
transitions. We already know how to get the parameter and label specific
transitions for each replica. What we really need is a transition
to exchange replicas, say the replicas at inverse temperature $\beta_{i}$
and $\beta_{i+1}$, such that the detailed balance is preserved (Sec.
\ref{par:Detailed-balance-definition}):\begin{eqnarray*}
\pi_{ee}\left(\theta_{1},C_{1},\ldots,\theta_{i},C_{i},\theta_{i+1},C_{i+1},\ldots,\theta_{R},C_{R}\right)T_{i,i+1}\left(\theta_{i+1},C_{i+1},\theta_{i},C_{i}\mid\theta_{i},C_{i},\theta_{i+1},C_{i+1}\right) & =\\
\pi_{ee}\left(\theta_{1},C_{1},\ldots,\theta_{i+1},C_{i+1},\theta_{i},C_{i},\ldots,\theta_{R},C_{R}\right)T_{i,i+1}\left(\theta_{i},C_{i},\theta_{i+1},C_{i+1}\mid\theta_{i+1},C_{i+1},\theta_{i},C_{i}\right)\end{eqnarray*}

which leads to:\begin{eqnarray*}
\frac{T_{i,i+1}\left(\theta_{i},C_{i},\theta_{i+1},C_{i+1}\mid\theta_{i+1},C_{i+1},\theta_{i},C_{i}\right)}{T_{i,i+1}\left(\theta_{i+1},C_{i+1},\theta_{i},C_{i}\mid\theta_{i},C_{i},\theta_{i+1},C_{i+1}\right)} & = & \frac{\pi_{ee}\left(\theta_{1},C_{1},\ldots,\theta_{i},C_{i},\theta_{i+1},C_{i+1},\ldots,\theta_{R},C_{R}\right)}{\pi_{ee}\left(\theta_{1},C_{1},\ldots,\theta_{i+1},C_{i+1},\theta_{i},C_{i},\ldots,\theta_{R},C_{R}\right)}\\
 & = & \frac{\pi_{post,\beta_{i}}\left(\theta_{i},C_{i}\right)\cdot\pi_{post,\beta_{i+1}}\left(\theta_{i+1},C_{i+1}\right)}{\pi_{post,\beta_{i}}\left(\theta_{i+1},C_{i+1}\right)\cdot\pi_{post,\beta_{i+1}}\left(\theta_{i},C_{i}\right)}\\
 & = & \exp\left[-\left(\beta_{i}-\beta_{i+1}\right)\cdot\left(E\left(\theta_{i},C_{i}\right)-E\left(\theta_{i+1},C_{i+1}\right)\right)\right]\end{eqnarray*}

Again we write $T_{i,i+1}$ as a product of a proposal transition
kernel and an acceptance probability. Here we have already explicitly
chosen a deterministic proposal (we only propose transitions between
replicas at neighboring inverse temperatures) which gives us:\[
\frac{A_{i,i+1}\left(\theta_{i},C_{i},\theta_{i+1},C_{i+1}\mid\theta_{i+1},C_{i+1},\theta_{i},C_{i}\right)}{A_{i,i+1}\left(\theta_{i+1},C_{i+1},\theta_{i},C_{i}\mid\theta_{i},C_{i},\theta_{i+1},C_{i+1}\right)}=\exp\left[-\left(\beta_{i}-\beta_{i+1}\right)\cdot\left(E\left(\theta_{i},C_{i}\right)-E\left(\theta_{i+1},C_{i+1}\right)\right)\right]\]

It is therefore enough to take:\begin{equation}
A_{i,i+1}\left(\theta_{i},C_{i},\theta_{i+1},C_{i+1}\mid\theta_{i+1},C_{i+1},\theta_{i},C_{i}\right)=\min\left\{ 1,\exp\left[-\left(\beta_{i}-\beta_{i+1}\right)\cdot\left(E\left(\theta_{i},C_{i}\right)-E\left(\theta_{i+1},C_{i+1}\right)\right)\right]\right\} \label{eq:RE-Acceptance-proba}\end{equation}

The reader sees that if the state of the {}``hot'' replica ($\beta_{i+1}$)
has a lower energy ($E\left(\theta_{i},C_{i}\right)$) than the {}``cold''
one, the proposed exchange is always accepted. The exchange can pictorially
be seen as cooling down the hot replica and warming up the cold one.
Fundamentally this process amounts to make the replica which is at
the beginning \emph{and} at the end of the replica exchange transition
at the cold temperature to jump from one local minimum ($\theta_{i+1},C_{i+1}$)
to another one ($\theta_{i},C_{i}$). That is precisely what we were
looking for. The fact that we can as well accept unfavorable exchanges
(\emph{i.e}., raising the energy of the {}``cold'' replica and decreasing
the one of the {}``hot'' replica) is the price we have to pay for
our algorithm to generate samples from the proper posterior (we are
not doing optimization here).

In order for the replica exchange transition to work well we need
to be careful with our choice of inverse temperatures. The typical
energy of a replica (\emph{i.e}., its expected energy) increases when
$\beta$ decreases (Fig. \ref{cap:REM-TEST}A). We will therefore
typically have a positive energy difference: $\Delta E=E_{hot}-E_{cold}>0$
between the replicas at low and high $\beta$ before the exchange.
That implies that the acceptance ratio (Eq. \ref{eq:RE-Acceptance-proba})
for the replica exchange will be typically smaller than 1. Obviously,
if it becomes too small, exchanges will practically never be accepted.
To avoid this situation we need to choose our inverse temperatures
such that the typical product: $\Delta\beta\cdot\Delta E$, where
$\Delta\beta=\beta_{cold}-\beta_{hot}$, is close enough to zero \cite{Neal1994,HukushimaNemoto1996,Iba2001}.
In practice we used pre-runs with an a priori too large number of
$\beta$s, checked the resulting energy histograms and kept enough
inverse temperatures to have some overlap between successive histograms
(Fig. \ref{cap:REM-TEST}B).

In Sec. \ref{cap:Simple Data} we will perform replica exchange transitions
between each pair $\beta_{i}$, $\beta_{i+1}$ with an even, respectively
odd, \emph{i} at the end of each even, respectively odd, MC step.
With this replica exchange scheme, each MC time step will therefore
be composed of a complete parameter and label transition for each
replica, followed by a replica exchange transition. This scheme corresponds
to the one described by Hukushima and Nemoto \cite{HukushimaNemoto1996}.
A rather detailed account of the REM can be found in Mitsutake et
al \cite{Mitsutake2001}. Variations on this scheme do exist \cite{Neal1994,CeleuxEtAl2000}.

\section{An Example from a simulated data set\label{sec:An-Example}}

We will illustrate the performances of the algorithm with a simple
simulated data set. The data are supposed to come from 3 neurons which
are exactly described by our underlying model. Such an illustration
has in our opinion several advantages. Being simple it helps the reader
to concentrate on the inner working of the algorithm. Because the
data correspond to the underlying model hypothesis, our implementation
of the MCMC method should give us back the parameters used to simulate
the data, we are therefore performing here a simple and \emph{necessary}
test of our code. The data do moreover exhibit features (strong cluster
overlap, Fig. \ref{cap:Simple Data}B,C) which would make them unusable
by other algorithms. A presentation of the algorithm performances
with a much worse data set can be found in \cite{PouzatEtAl2004}.

\subsection{Data properties\label{sub:Data-properties}}

\begin{figure}
\begin{center}

\includegraphics[%
  scale=0.5,
  angle=270]{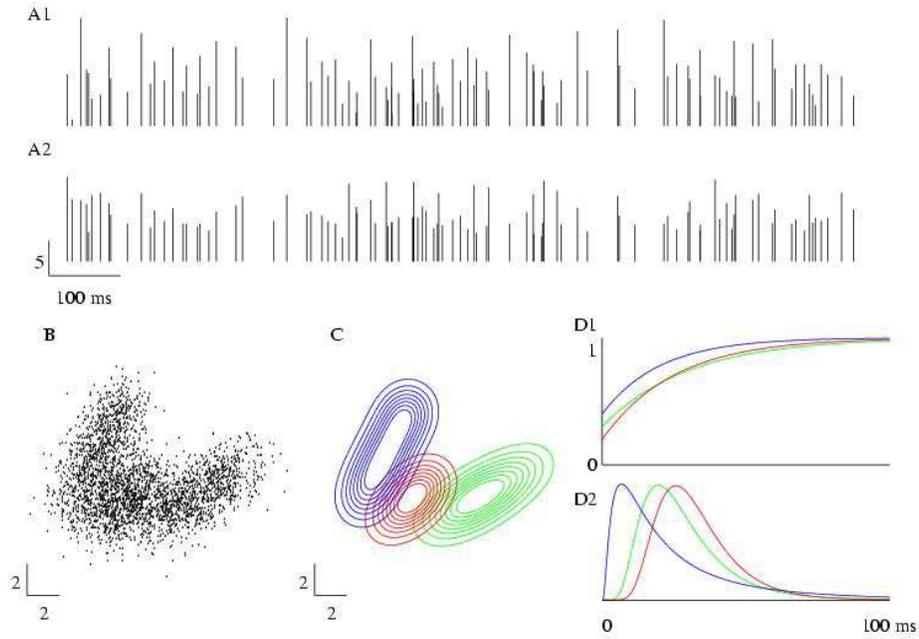}

\end{center}

\caption{Simulated data with 3 active neurons recorded on a stereode (2 recording
sites). 30 s of data were simulated resulting in 2966 spikes. A, An
epoch with 100 spikes on site 1 (A1) and 2 (A2). Vertical scale bar
in units of noise SD. B, Wilson plot of the 2966 spikes: the amplitude
of each spike on site 2 is plotted against its amplitude on site 1.
The scale bars corner is at the origin (0,0). C, Expected iso-density
plots for each of the three neurons: neuron 1 (green), neuron 2 (red)
and neuron 3 (blue). D1, Normalized spike amplitude \emph{vs} ISI
for each neuron (normalization done with respect to the maximal amplitude).
Horizontal scale, same as D2. D2, ISI density of each of the 3 neurons
(peak value of blue density: 35 Hz). \label{cap:Simple Data}}
\end{figure}

The parameters used to simulate the three neurons are given in Table
\ref{cap:Data-Parameters}. 30 seconds of data were simulated giving
a total of 2966 spikes. The raw data, spike amplitude \emph{vs} time
on the two recording sites are illustrated on Fig. \ref{cap:Simple Data}A1
\& \ref{cap:Simple Data}A2. Fig. \ref{cap:Simple Data}B is a {}``Wilson
plot'' of the entire sample. Notice the \emph{strong} overlap of
points (spikes) arising from the 3 different neurons. Fig. \ref{cap:Simple Data}C
shows the theoretical iso-density contours for the clusters generated
by each of the 3 neurons. Neuron 1 in Table \ref{cap:Data-Parameters}
is green on the figure, neuron 2 is red and neuron 3 is blue. The
reader can see that roughly 50\% of the spikes generated by neuron
2 should fall in regions were neuron 1 or neuron 3 will also generate
spikes. The theoretical densities associated with each neuron (cluster)
\emph{are not} 2-dimensional Gaussian. This can be most clearly seen
for neuron 3 (blue iso-density contours) which has a {}``flat''
summit. None of these densities is symmetrical with respect to its
maximum along its principal axis (which is the axis going through
the graph's origin and the neuron density maximum). Fig. \ref{cap:Simple Data}D1
represents the amplitude dynamics of each of the three neurons, while
Fig. \ref{cap:Simple Data}D2 displays their respective ISI densities.

\begin{table}
\begin{center}

\begin{tabular}{cccc}
&
neuron 1&
neuron 2&
neuron 3\tabularnewline
$P_{1},P_{2}$&
15,9&
8,8&
6,12\tabularnewline
$\delta$&
0.7&
0.8&
0.6\tabularnewline
$\lambda$&
33.33&
40&
50\tabularnewline
$s$&
25&
30&
18\tabularnewline
$f$&
0.5&
0.4&
1.0\tabularnewline
$\left\langle isi\right\rangle $&
28.3&
32.5&
29.7\tabularnewline
\#&
1060&
923&
983\tabularnewline
\end{tabular}

\end{center}

\caption{\label{cap:Data-Parameters}Parameters used to simulate the neurons.
The maximal peak amplitude values ($P_{i}$) are given in units of
noise SD. The scale parameters (\emph{s}) and mean \emph{isi} ($\left\langle isi\right\rangle $)
are given in ms. The bottom row indicates the number of events from
each neuron. The correspondence between neuron number and color on
Fig. \ref{cap:Simple Data}: 1, green, 2, red, 3, blue.}
\end{table}

\subsection{Algorithm dynamics and parameters estimates without the REM\label{sub:Algorithm-dynamics-NOREM}}

\subsubsection{Initialization}

The algorithm being iterative we have to start it somewhere and we
will use here a somewhat {}``brute force'' initialization. We choose
randomly with a uniform probability $\frac{1}{N}$ as many actual
events as neurons in the model (\emph{K}=3). That gives us our initial
guesses for the $P_{q,i}$. $\delta$ is set to $\delta_{min}=0$
for each neuron. All the other parameters are randomly drawn from
their prior distribution (Sec. \ref{sec:Choice-of-the-priors}). The
initial configuration is generated by labeling each individual spike
with one of the \emph{K} possible labels with a probability $1/K$
for each label (this is the $\beta=0$ initial condition used in statistical
physics).

\subsubsection{Energy evolution}

\begin{figure}
\begin{center}

\includegraphics[%
  scale=0.4,
  angle=270]{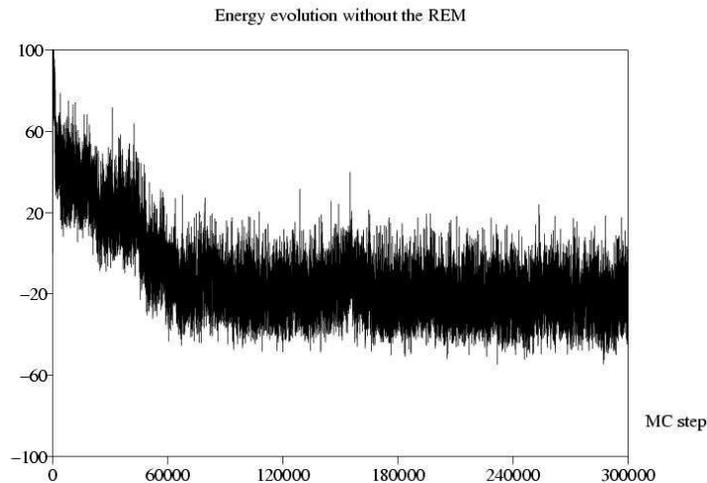}

\end{center}

\caption{\label{cap:NRJ-EVOL-NOREM}Energy evolution without the REM. Notice
the step like decreases in the early part.}
\end{figure}

At each MC step, a new label is proposed and accepted or rejected
for each spike and a new value is proposed and accepted or rejected
for each of the (18) model parameters (Eq. \ref{eq:Complete-MC-Step}).
Our initial state is very likely to have a very low posterior probability.
We therefore expect our system to take some time to relax from the
highly disordered state in which we have initialized it to the (hopefully)
ordered typical states. As explained in Sec. \ref{sub:Initialization-bias},
we can (must) monitor the evolution of {}``system features''. The
most reliable we have found until now is the system's energy as shown
on Fig. \ref{cap:NRJ-EVOL-NOREM}. Here one sees an {}``early''
evolution (first 60000 MC steps) followed by a much slower one which
\emph{looks like} a stationary regime during the last 100000 steps%
\footnote{The time required to perform such simulations depends on the number
of sampling points used for the piece-wise linear proposals of the
amplitude parameters (Sec. \ref{sub:Generation-of-the-Amplitude-Para}).
During the first 2000 steps we use 101 regularly spaced sampling points.
We then need less than 5' to perform 1000 steps. After that, we reposition
the sampling points and use only 13 of them and the time required
to perform 1000 steps falls to 50 s (Pentium IV computer at 3.06 GHz).%
}.

\subsubsection{Model parameters evolution}

\begin{figure}
\begin{center}

\includegraphics[%
  scale=0.4,
  angle=270]{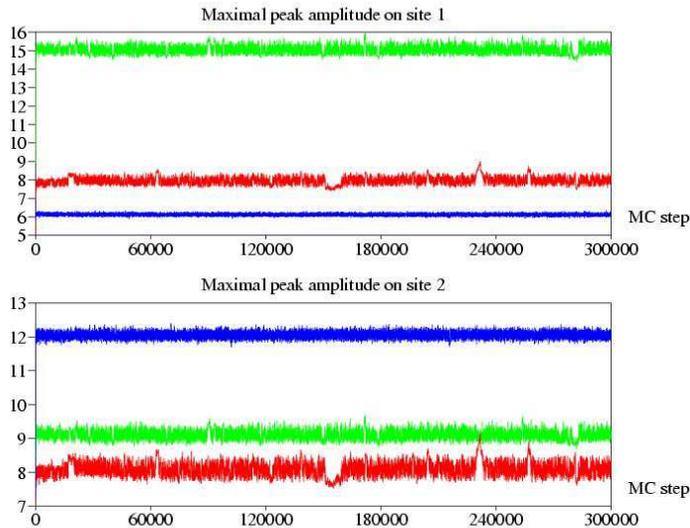}

\end{center}

\caption{\label{cap:PEAK-EVOL-NOREM}Evolution of the maximal peak amplitude
on the two recording sites for the three neurons. Notice that an apparently
stationary regime is reached quickly for those parameters, while the
energy (Fig. \ref{cap:NRJ-EVOL-NOREM}) is still slowly decreasing.}
\end{figure}

We can monitor as well the evolution of models parameters like the
maximal peak amplitudes of the neurons%
\footnote{We monitor in practice the evolution of every model parameter. We
show here only the evolution of the maximal peak amplitude to keep
this chapter length within bounds.%
} as shown on Fig. \ref{cap:PEAK-EVOL-NOREM}. The interesting, and
typical, observation here is that the model parameters reach (apparent)
stationarity much earlier than the energy (compare the first 60000
steps on Fig. \ref{cap:NRJ-EVOL-NOREM} \& \ref{cap:PEAK-EVOL-NOREM}).
That means that most of the slow relaxation of the energy has a configurational
origin. In other words, it comes from spike labels re-arrangements.

\subsubsection{Model parameters estimates}

\begin{figure}
\begin{center}

\includegraphics[%
  scale=0.5,
  angle=270]{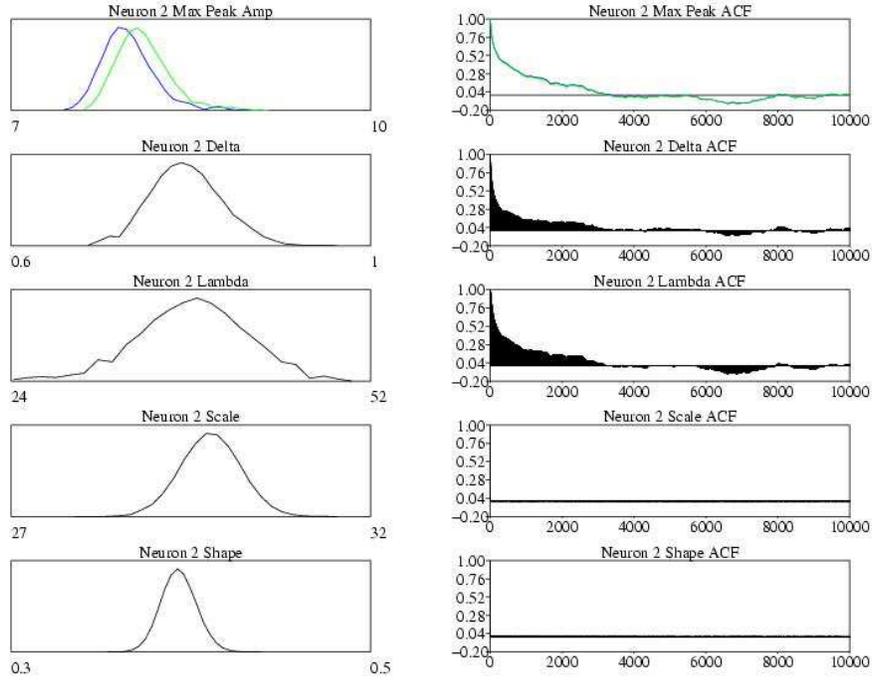}

\end{center}

\caption{\label{cap:N2-PE-NOREM}Marginal posterior estimates for the parameters
of neuron 2. The left part shown the estimated density of each of
the six parameters (for the peak amplitudes, the first site corresponds
to the blue curve and the second to the green curve). The right side
shows the corresponding ACFs (Sec. \ref{sub:Autocorrelation-functions-and}
and Eq. \ref{eq:Autocorrelation}), the abscissa is the lag in MC
steps.}
\end{figure}

Based on the energy trace (Fig. \ref{cap:NRJ-EVOL-NOREM}) and on
the evolution of the model parameters (\emph{e.g.,} Fig. \ref{cap:PEAK-EVOL-NOREM})
we could reasonably decide to keep the last 100000 MC steps for {}``measurements''
and estimate the posterior (marginal) densities of our model parameters
from these steps. We would then get for neuron 2 what's shown on Fig.
\ref{cap:N2-PE-NOREM}.%
The striking feature of this figure is the time taken by the ACFs
of the amplitude parameters ($P_{1},P_{2},\delta,\lambda$) to reach
0: 4000 steps. With the present data set we observe a similar behavior
for the amplitude parameters of the first neuron but not for the third.
That explains the un-acceptably low precision on the parameters estimates
reported in Table \ref{cap:Para-Estimates-NOREM} for neurons 1 \&
2.

\begin{table}
\begin{center}

\begin{tabular}{cccc}
&
\textbf{neuron 1}&
\textbf{neuron 2}&
\textbf{neuron 3}\tabularnewline
$\overline{P_{1}}$&
$15\pm7$&
$8\pm7$&
$6.1\pm0.2$\tabularnewline
&
(1071)&
(1052)&
(16)\tabularnewline
$\overline{P_{2}}$&
$9\pm4$&
$8\pm6$&
$12.1\pm0.5$\tabularnewline
&
(945)&
(930)&
(27)\tabularnewline
$\overline{\delta}$&
$0.7\pm0.3$&
$0.8\pm1.2$&
$0.58\pm0.05$\tabularnewline
&
(586)&
(941)&
(32)\tabularnewline
$\overline{\lambda}$&
$33\pm64$&
$39\pm137$&
$47\pm13$\tabularnewline
&
(1250)&
(1043)&
(45)\tabularnewline
$\overline{s}$&
$24.9\pm0.1$&
$29.8\pm0.1$&
$18.3\pm0.5$\tabularnewline
&
(6.5)&
(7)&
(1)\tabularnewline
$\overline{f}$&
$0.51\pm0.05$&
$0.40\pm0.05$&
$1.01\pm0.03$\tabularnewline
&
(6.5)&
(7)&
(1)\tabularnewline
\end{tabular}

\end{center}

\caption{\label{cap:Para-Estimates-NOREM}Estimated parameters values using
the last $10^{5}$ MC steps (among $3\cdot10^{5}$). The SDs are autocorrelation
corrected (Sec. \ref{sub:Autocorrelation-functions-and}). The integrated
autocorrelation time (Eq. \ref{eq:Autoco-time}) is given below each
parameter between {}``()''.}
\end{table}

\subsection{Algorithm dynamics and parameters estimates with the REM\label{sub:Algorithm-dynamics-REM}}

\begin{figure}
\begin{center}

\includegraphics[%
  scale=0.4,
  angle=270]{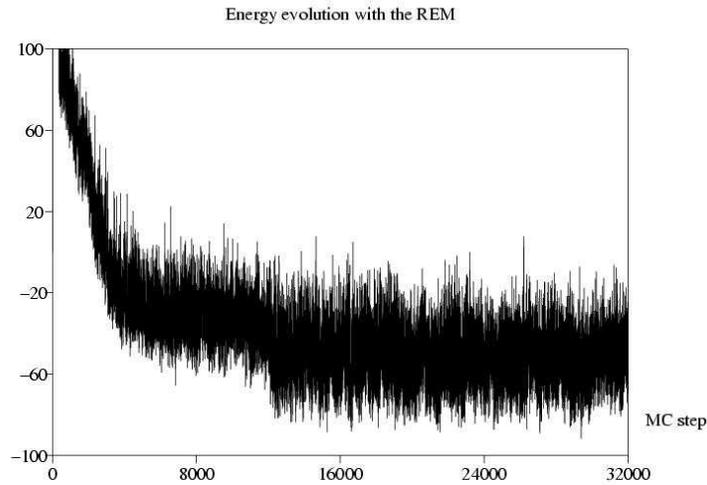}

\end{center}

\caption{\label{cap:NRJ-EVOL-REM}Energy evolution with the REM. The ordinate
scale is the same as in Fig. \ref{cap:NRJ-EVOL-NOREM}.}
\end{figure}

The different behaviors of the model parameters, which relax to (apparent)
equilibrium relatively fast ($2\cdot10^{4}$ to $3\cdot10^{4}$ MC
steps) and of the energy which relaxes more slowly ($6\cdot10^{4}$
MC steps) suggests that the configuration dynamics is slower than
the model parameters dynamics. It could therefore be worth considering
a modification of our basic algorithm which could speed up the configuration
(and if we are lucky the model parameters) dynamics like the Replica
Exchange Method (Sec. \ref{sec:Slow-relaxation-and-REM}). To illustrate
the effect of the REM on our simple data set we have taken the first
2000 MC steps of the previous section and restarted%
\footnote{When we introduce new $\beta$ values from one run to the next, the
initial state of the replica with the new $\beta$ values is set to
the final state of the replica with the closest $\beta$ in the previous
run. That is, in our example, at step 2001 all the replicas are in
the same state (same parameters values, same configuration).%
} the algorithm augmenting our inverse temperature space from $\beta\in\left\{ 1\right\} $
to $\beta\in\left\{ 1,0.95,0.9,0.85,0.8,0.75,0.7,0.65,0.6,0.55,0.5\right\} $.
The resulting energy trace for $\beta=1$ and 30000 additional steps
is shown on Fig. \ref{cap:NRJ-EVOL-REM}. The striking feature here
is that the energy reaches after roughly $1.2\cdot10^{4}$ steps a
\emph{lower} energy than after $30\cdot10^{4}$ steps without the
REM. Our REM scheme with 11 inverse temperatures implies that each
step requires 11 times more CPU time than one step with a single inverse
temperature%
\footnote{The computational overhead required accept or reject the proposed
replica exchanges (Eq. \ref{eq:RE-Acceptance-proba}) is negligible
compared to the time required to update every spike label and every
model parameter.%
}. From the energy evolution view point the REM is clearly more efficient
than a single long simulation at $\beta=1$. A more striking illustration
of the REM effect on energy relaxation (on an other data set) can
be found in \cite{PouzatEtAl2004}.

\subsubsection{Making sure the REM {}``works''\label{sub:Making-sure}}

\begin{figure}
\begin{center}

\includegraphics[%
  scale=0.4,
  angle=270]{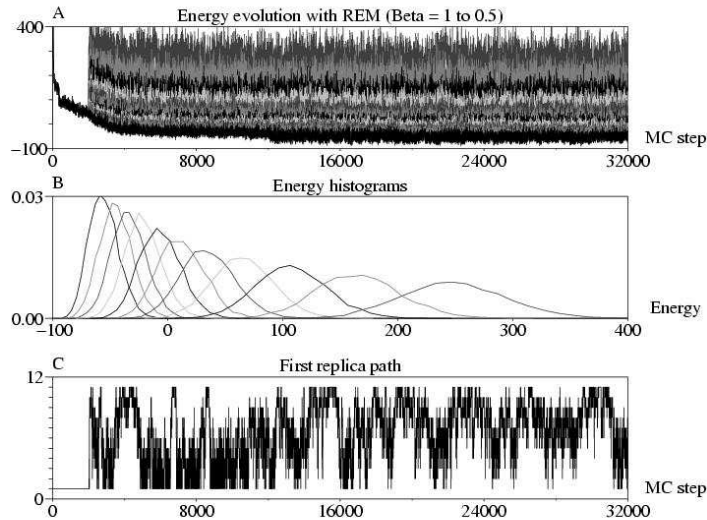}

\end{center}

\caption{\label{cap:REM-TEST}Test of the REM. A, Energy evolution for several
$\beta$ values (see text) during successive runs. 1 replica is used
between steps 1 and 2000, 11 replicas are used between steps 2001
and 32000. B, Energy histograms at each $\beta$ value (histograms
computed from the 10000 last steps with 25 bins): the left histogram
correspond to $\beta=1$, the right one to $\beta=0.5$. C, Path of
the first replica in the temperature space. The ordinate corresponds
to the $\beta$ index (1 corresponds to $\beta=1$ and 11 to $\beta=0.5$
).}
\end{figure}

As explained in Sec. \ref{sec:Slow-relaxation-and-REM} the REM will
work efficiently only if the energies explored by replicas at adjacent
$\beta$ exhibit enough overlap. Fig. \ref{cap:REM-TEST} illustrates
how the efficiency of the {}``replica exchange'' can be empirically
assessed \cite{Hansman1997,YandePablo1999,FalcioniDeem1999,Mitsutake2001}.
Energy traces at different $\beta$ values are displayed on Fig.\ref{cap:REM-TEST}A.
The overlap between adjacent energy traces is already clear. The lowest
trace ($\beta=1$) is the trace of Fig. \ref{cap:NRJ-EVOL-REM}. Fig.
\ref{cap:REM-TEST}B shows the energy histograms for the different
$\beta$. Adjacent histograms clearly exhibit a significant overlap.
A pictorial way to think of the REM is to imagine that several {}``boxes''
at different pre-set inverse-temperatures are used and that there
is one and only one replica per box at each step. After each MC step,
the algorithm proposes to exchange the replicas located in neighboring
boxes (neighboring in the sense of their inverse-temperatures) and
this proposition can be accepted or rejected (Eq. \ref{eq:RE-Acceptance-proba}).
Then if the exchange dynamics works properly one should see each replica
visit all the boxes during the MC run. More precisely each replica
should perform a random walk on the available boxes. Fig. \ref{cap:REM-TEST}C
shows the random walk performed by the replica which start at $\beta=1$
at step 2001. The ordinate corresponds to the box index (see legend).
Between steps 2001 and 32000, the replica travels through the entire
inverse temperature range. 

One of the shortcomings of the REM is that it requires more $\beta$
to be used (for a given range) as the number of spikes in the data
set increases because the width of the energy histograms (Fig. \ref{cap:REM-TEST}B)
is inversely proportional to the square root of the number \emph{N}
of events in the sample \cite{HukushimaNemoto1996,Hansman1997,Iba2001}.
The necessary number of $\beta$ grows therefore as $\sqrt{N}$. The
computation time per replica grows, moreover, linearly with \emph{N}.
We therefore end up with a computation time of the REM growing like:
$N^{1.5}$.

\subsubsection{Posterior estimates with the REM\label{sub:Posterior-estimates-REM}}

We do not present here the Equivalent of Fig. \ref{cap:PEAK-EVOL-NOREM}
with the REM because the two figures would look too similar. The best
way to see the effect of the REM on the model parameters dynamics
is to look at the empirical \emph{integrated autocorrelation time}
(IAT) for each parameter as show in Table \ref{cap:Para-Estimates-REM}.
When we compare these values with the ones obtained with our basic
algorithm (without the REM), we see for instance that longest IAT
is now 110 steps (parameter $\lambda$ of the second neuron) while
the longest IAT without the REM was 1250 steps (parameter $\lambda$
of the first neuron, Table \ref{cap:Para-Estimates-NOREM}). We therefore
get \emph{much better} \emph{parameters estimates although the absolute
sample size we use $10^{4}$ is 10 times smaller than the one used
previously}. This means that in this simple setting we were able without
any fine tunning (we could have in fact used fewer inverse temperatures,
we have not optimized the number of steps between replica exchange
attempts) to get a much more efficient algorithm (in term of statistical
efficiency and of relaxation to equilibrium) without extra computational
cost.

\begin{table}
\begin{center}

\begin{tabular}{cccc}
&
\textbf{neuron 1}&
\textbf{neuron 2}&
\textbf{neuron 3}\tabularnewline
$\overline{P_{1}}$&
$15\pm2$&
$8\pm2$&
$6.1\pm0.1$\tabularnewline
&
(62)&
(91)&
(12)\tabularnewline
$\overline{P_{2}}$&
$9\pm1$&
$8\pm2$&
$12.1\pm0.3$\tabularnewline
&
(55)&
(93)&
(14)\tabularnewline
$\overline{\delta}$&
$0.70\pm0.07$&
$0.8\pm0.3$&
$0.58\pm0.02$\tabularnewline
&
(25)&
(93)&
(9)\tabularnewline
$\overline{\lambda}$&
$34\pm19$&
$40\pm38$&
$46\pm7$\tabularnewline
&
(60)&
(110)&
(20)\tabularnewline
$\overline{s}$&
$24.9\pm0.6$&
$30.0\pm0.4$&
$18.3\pm0.7$\tabularnewline
&
(2)&
(1.5)&
(2.5)\tabularnewline
$\overline{f}$&
$0.51\pm0.02$&
$0.40\pm0.01$&
$1.01\pm0.03$\tabularnewline
&
(2)&
(1.5)&
(2.5)\tabularnewline
\end{tabular}

\end{center}

\caption{\label{cap:Para-Estimates-REM}Estimated parameters values using
the last $10^{4}$ MC steps (among $6\cdot10^{4}$). The SDs are autocorrelation
corrected (Sec. \ref{sub:Autocorrelation-functions-and}). The integrated
autocorrelation time (Eq. \ref{eq:Autoco-time}) is given below each
parameter between {}``()''.}
\end{table}

\subsubsection{Configuration estimate}

As a quick way to compare estimated and actual configurations we can
introduce what we will (abusively) call the {}``most likely'' configuration
estimate. First, we estimate the probability for each spike to originate
from each neuron. For instance, if we discard the first $N_{D}=22000$
on a total of $N_{T}=32000$ steps we have for the probability of
the 100th spike to have been generated by the second neuron:\begin{equation}
Pr(l_{100}=2\mid Y)\approx\frac{1}{10000}\,\cdot\sum_{t=22000}^{32000}\mathcal{{I}}_{2}\left[l_{100}^{(t)}\right]\label{eq:Proba-Label}\end{equation}

where $\mathcal{{I}}_{q}$ is the \emph{indicator} function defined
Eq. \ref{eq:Indicator Function}.

\begin{figure}
\begin{center}

\includegraphics[%
  scale=0.4,
  angle=270]{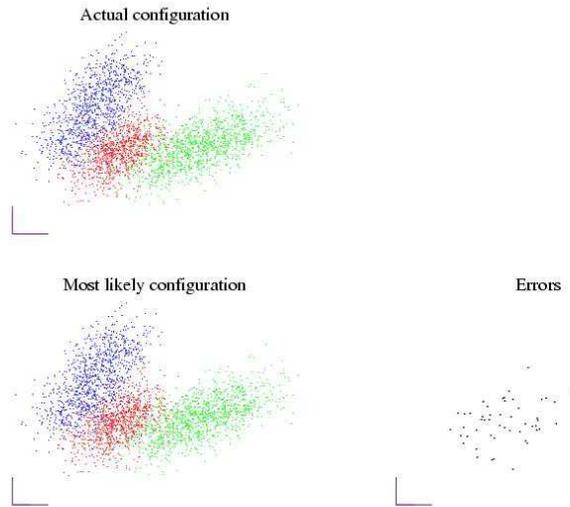}

\end{center}

\caption{\label{cap:WILSON-COMP-REM}Comparison between the actual and estimated
configurations. Top, the Wilson plot of Fig. \ref{cap:Simple Data}B
color coded to show the neuron of origin of each spike. Bottom left,
the most likely configuration (see text) estimated from the last 10000
steps with the REM. Bottom right, the 50 errors (among 2966 events).}
\end{figure}

We then {}``force'' the spike label to its most likely value. That
is, if we have: $Pr(l_{100}=1\mid Y)=0.05$, $Pr(l_{100}=2\mid Y)=0.95$
and $Pr(l_{100}=3\mid Y)=0$, we force $l_{100}$ to 2. We proceed
in that way with each spike to get our {}``most likely'' configuration.
Fig. \ref{cap:WILSON-COMP-REM} shows what we get using the last $10\cdot10^{3}$
MC steps of our run using the REM. The actual configuration is shown
too, as well as the \textbf{50} misclassified events. We therefore
get \emph{1.7\% misclassified spikes with our procedure}. %
 We leave as an excercise to the reader to check that using an exact
knowledge of the model parameters and the theoretical 2-dimensional
Wilson plot densities one can get from them (Fig. \ref{cap:Simple Data}C)
we would get roughly 9\% of the spikes misclassified. That would be
the optimal classification we could generate using only the amplitude
information. Our capacity to go beyond these 9\% clearly stems from
our inclusion of the ISI density in our data generation model. Using
the last $10\cdot10^{3}$ MC steps of our run without the REM we generate
63 misclassifications, while we get {}``only'' 59 of them with the
last $100\cdot10^{3}$ steps%
\footnote{This difference in the number of misclassification explains the energy
difference observed in the two runs (Fig. \ref{cap:NRJ-EVOL-NOREM}
\& \ref{cap:NRJ-EVOL-REM}). It means clearly that the run without
the REM has not reached equilibrium.%
}.

\subsubsection{A more detailed illustration of the REM dynamics.}

\begin{figure}
\begin{center}

\includegraphics[%
  scale=0.4,
  angle=270]{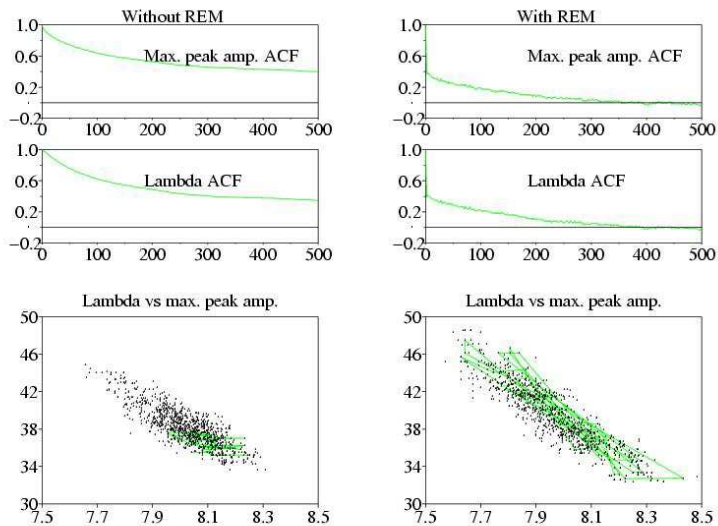}

\end{center}

\caption{\label{cap:WHY-REM-WORKS}REM and parameters auto-correlation functions.
The ACFs of the maximal peak amplitude of neuron 2 on site 1 ($P_{2,1}$
) are shown together with the ACFs of the parameter $\lambda$ of
neuron 2. The ACFs without the REM (left side) were computed from
the 100000 last iterations (Fig. \ref{cap:NRJ-EVOL-NOREM}). The ACFs
with the REM (right side) were computed from the 10000 last iterations
(Fig. \ref{cap:NRJ-EVOL-REM}). The ACFs with REM clearly fall faster
than the ones without. The two graphs at the bottom give a qualitative
explanation of why it is so. On each graph the last 1000 generated
values for $P_{2,1}$ (abscissa) and $\lambda_{1}$ (ordinate) are
shown (black dots). 50 values generated \emph{successively} are linked
together by the broken lines (green). The size of the jumps is clearly
larger with REM than without.}
\end{figure}

When the posterior density one wants to explore is non isotropic a
problem arise with the use of an MH algorithm based on component specific
transitions. In such cases, the transitions proposed by the MH algorithm
are not optimal in the sense that only {}``small'' changes of the
present parameters values are likely to be accepted. These {}``small''
changes will lead to a {}``slow'' exploration of the posterior density
and therefore to {}``long'' autocorrelation times. This is illustrated
for our MH algorithm without the REM on the left side of Fig. \ref{cap:WHY-REM-WORKS}.
Here we show only the situation in the plane defined by the maximal
peak amplitude on site 1 and the parameter $\lambda$ of neuron 1
($P_{2,1},\lambda_{1}$) but the reader should imagine that the same
holds in the 4 dimensional spaces defined by the 4 amplitude parameters
of each neurons. The ACFs (Eq. \ref{eq:Autocorrelation}) of the two
parameters are shown. The last 1000 values of the parameters generated
by the MH algorithm without the REM with the last 50 of them linked
together by the broken line are shown. The movement of the {}``system''
in its state space has clear Brownian motion features: small steps,
random direction change from one step to the next, resulting in a
{}``slow'' exploration of the posterior density. If we now look
at the system's dynamics using the REM, we see that the ACFs fall
to zero much faster (right side of Fig. \ref{cap:WHY-REM-WORKS}).
This is easily understood by observing successive steps of the systems
in the corresponding plane ($\lambda$ vs $P_{2,1}$ graph, bottom).
Large steps going almost from one end to the other of the marginal
posterior density are now observed, meaning that the resulting Markov
chain explores very efficiently the posterior density we want to study.
This is clearly due to the fact that the high temperature replicas
can make much larger steps with a still reasonably high acceptance
probability. The replica exchange dynamics (cooling down {}``hot''
replicas and warming up {}``cold'' ones) does the rest of the job.
This increased statistical efficiency ($\tau_{autoco}$ gets smaller)
does in fact compensate for the increased computational cost ($\tau_{cpu}$
gets larger) of the REM (compare the integrated autocorrelation times
in Table \ref{cap:Para-Estimates-NOREM} \& \ref{cap:Para-Estimates-REM}).

\section{Conclusions}

We have given a detailed presentation of a simple data generation
model for extracellular recording. Although simple this model is already
much more sophisticated than the ones previously used \cite{Lewicki1994,Sahani1999,PouzatEtAl2002,NugyenEtAl2003},
but this sophistication comes at a price: the usual algorithms presented
in the spike-sorting literature cannot be used. To work with our model
we had to find an analogy between our problem and the Potts spin glass
model studied in statistical physics. This analogy allowed us to get
a Dynamic MC/MCMC algorithm with which model parameters and configurations
could be estimated \emph{with confidence intervals}.

We have not explained here how one of the critical questions of spike-sorting,
the number of neuron contributing to the data, can be tackled. We
have recently described in \cite{PouzatEtAl2004} how this can be
done with an ad-hoc method. The next step is clearly a Bayesian solution
to this problem which fundamentally implies estimating the normalizing
constant \emph{Z} of Eq. \ref{eq:Posterior PDF} and \ref{eq:Normalizing Constant}
(which is the probability of the data). Nguyen et al \cite{NugyenEtAl2003}
have already given a solution for a simple data generation model.
This task of estimating \emph{Z} could seem daunting but is luckily
not. Dynamic MC and MCMC methods have been used for a long time which
means that other people in other fields already had that problem.
Among the proposed solutions, what physicists call thermodynamic integration
\cite{LandauBinder2000,FrenkelSmit2002} seems very attractive because
it is known to be robust and it requires simulations to be performed
between $\beta$ = 1 and $\beta$ = 0. And that's precisely what we
are already (partially) doing with the REM. Of course the estimation
of \emph{Z} is only half of the problem. The prior distribution on
the number of neurons in the model has to be set properly as well.
Increasing the number of neurons will always lead to a decrease in
energy which will give larger \emph{Z} values (the derivative of the
log of \emph{Z} being proportional to the opposite of the energy \cite{FrenkelSmit2002}),
we will therefore have to compensate for this systematic \emph{Z}
increase with the prior distribution.

Finally, although our method requires fairly long computation to be
carried out, we hope we have convinced our reader that the presence
of strong overlaps on Wilson plots does not necessarily precludes
good sorting.

\section{Exercises solutions}

\subsection{Cholesky decomposition (or factorization)\label{sub:Cholesky-decomposition}}

Write $a_{i,j}$ the elements of \emph{A} and $\sigma_{i,j}$ the
elements of $\Gamma$, compute:\[
AA^{T}=B\]

with: $b_{i,j}=\sum_{k=1}^{D}a_{i,k}a_{j,k}$ and identify individual
terms, that is:\[
a_{1,1}=\sqrt{\sigma_{1,1}},\]
where the fact that \emph{A} is lower triangular has been used.\[
a_{,j,1}=\frac{\sigma_{j,1}}{a_{1,1}}\:,\,2\leq j\leq D.\]

\[
a_{2,2}=\sqrt{\sigma_{2,2}-a_{2,1}^{2}}.\]

And keep going like that and you will find:\begin{eqnarray*}
for\; i & = & 1\; to\; D\\
for\; j & = & i+1\; to\; D\\
a_{i,i} & = & \sqrt{\sigma_{i,i}-\sum_{k=1}^{i-1}a_{i,k}^{2}}\\
a_{j,i} & = & \frac{\sigma_{j,i}-\sum_{k=1}^{i-1}a_{j,k}a_{i,k}}{a_{i,i}}\end{eqnarray*}

For details see \cite{BockKrischer1998}.

\subsection{Bayesian posterior densities for the parameters of a log-Normal \emph{pdf\label{sub:Bayesian-for-log-Normal}}}

\paragraph{Answer to question 1}

We have:\[
\pi\left(s\,\mid f,\mathcal{{D}}\right)\,\alpha\,\exp\left[-\frac{1}{2}\,\frac{1}{f^{2}}\,\sum_{j=1}^{N}\left(\ln i_{j}-\ln s\right)^{2}\right]\cdot\frac{1}{s_{max}-s_{min}}\cdot\mathcal{{I}}_{\left[s_{min},s_{max}\right]}\left(s\right).\]

If we introduce $\overline{\ln i}$ defined by Eq. \ref{eq:Mean Log ISI}
in the summed term we get:\begin{eqnarray*}
\sum_{j=1}^{N}\left(\ln i_{j}-\ln s\right)^{2} & = & \sum_{j=1}^{N}\left(\ln i_{j}-\overline{\ln i}+\overline{\ln i}-\ln s\right)^{2}\\
 & = & \sum_{j=1}^{N}\left(\ln i_{j}-\overline{\ln i}\right)^{2}+2\,\left(\overline{\ln i}-\ln s\right)\cdot\sum_{j=1}^{N}\left(\ln i_{j}-\overline{\ln i}\right)+N\,\left(\overline{\ln i}-\ln s\right)\end{eqnarray*}

The first term does not depend on \emph{s}, the second is zero, we
therefore have:\[
\pi\left(s\,\mid f,\mathcal{{D}}\right)\,\alpha\,\exp\left[-\frac{1}{2}\,\frac{N}{f^{2}}\,\left(\overline{\ln i}-\ln s\right)^{2}\right]\cdot\frac{1}{s_{max}-s_{min}}\cdot\mathcal{{I}}_{\left[s_{min},s_{max}\right]}\left(s\right).\]

\paragraph{Answer to question 2}

\begin{enumerate}
\item Generate $U\,\sim\, Norm\left(\overline{\ln i},\frac{f^{2}}{N}\right)$%
\footnote{This notation means that the random variable \emph{U} has a normal
distribution with mean: $\overline{\ln i}$ and variance: $\frac{f^{2}}{N}$
.%
}
\item if $s_{min}\leq\exp u\leq s_{max}$, then $S=\exp u$\\
otherwise, go back to 1
\end{enumerate}

\paragraph{Answer to question 3}

Using the first line of Answer 1 we get:\[
\pi\left(f\,\mid s,\mathcal{{D}}\right)\,\alpha\,\frac{1}{f^{N}}\,\exp\left[-\frac{1}{2}\frac{\sum_{j=1}^{N}\left(\ln i_{j}-\ln s\right)^{2}}{f^{2}}\right]\cdot\frac{1}{f_{max}-f_{min}}\cdot\mathcal{{I}}_{\left[f_{min},f_{max}\right]}\left(f\right).\]

Identification does the rest of the job.

\paragraph{Answer to question 4}

Using $\theta=\frac{1}{\omega}$ and the Jacobian of the change of
variable, you can easily show that if $\Omega\,\sim\, Gamma\left(\alpha,\beta\right)$
then $\Theta=\frac{1}{\Omega}\,\sim\, Inv-Gamma\left(\alpha,\beta\right)$.
The algorithm is simply:

\begin{enumerate}
\item Generate $\Omega\,\sim\, Gamma\left(\frac{N}{2}-1,\frac{1}{2}\,\sum_{j=1}^{N}\left(\ln i_{j}-\ln s\right)^{2}\right)$
\item if $f_{min}\leq\sqrt{\frac{1}{\omega}}\leq f_{max}$ , then $F=\sqrt{\frac{1}{\omega}}$\\
otherwise go back to 1.
\end{enumerate}

\subsection{Stochastic property of a Markov matrix\label{sub:Stochastic Property of MC}}

$X^{\left(t+1\right)}$ being a \emph{rv} defined on $\mathcal{{X}}$
we must have by definition:\[
Pr\left(X^{\left(t+1\right)}\subset\mathcal{{X}}\right)=1,\]

where $X^{\left(t+1\right)}\subset\mathcal{{X}}$ should be read as:
$X^{\left(t+1\right)}=x_{1}+x_{2}+\ldots+x_{\nu}$ (remember that
the $x_{i}$ are mutually exclusive events). We have therefore:\begin{eqnarray*}
Pr\left(X^{\left(t+1\right)}\subset\mathcal{{X}}\right) & = & \sum_{i=1}^{\nu}Pr\left(X^{\left(t+1\right)}=x_{i}\right)\\
 & = & \sum_{i=1}^{\nu}Pr\left(X^{\left(t+1\right)}=x_{i}\mid X^{\left(t\right)}\subset\mathcal{{X}}\right)Pr\left(X^{\left(t\right)}\subset\mathcal{{X}}\right)\\
 & = & \sum_{i=1}^{\nu}\sum_{j=1}^{\nu}Pr\left(X^{\left(t+1\right)}=x_{i}\mid X^{\left(t\right)}=x_{j}\right)\, Pr\left(X^{\left(t\right)}=x_{j}\right)\\
 & = & \sum_{j=1}^{\nu}Pr\left(X^{\left(t\right)}=x_{j}\right)\left[\sum_{i=1}^{\nu}Pr\left(X^{\left(t+1\right)}=x_{i}\mid X^{\left(t\right)}=x_{j}\right)\right]\end{eqnarray*}

Which implies:\[
\sum_{j=1}^{\nu}Pr\left(X^{\left(t\right)}=x_{j}\right)\left[\sum_{i=1}^{\nu}Pr\left(X^{\left(t+1\right)}=x_{i}\mid X^{\left(t\right)}=x_{j}\right)\right]=1\]

And that must be true for any $X^{\left(t\right)}$ which implies
that%
\footnote{If you're not convinced you can do it by \emph{absurdum} assuming
it's not the case and then find a $X^{\left(t\right)}$which violate
the equality.%
}:\[
\sum_{i=1}^{\nu}Pr\left(X^{\left(t+1\right)}=x_{i}\mid X^{\left(t\right)}=x_{j}\right)=1\]

\subsection{Detailed balance\label{sub:Detailed Balance}}

If $\forall x,y\,\in\,\mathcal{{X}},\:\pi\left(x\right)T\left(x,y\right)=\pi\left(y\right)T\left(y,x\right)$
then:\begin{eqnarray*}
\sum_{x\,\in\,\mathcal{{X}}}\pi\left(x\right)T\left(x,y\right) & = & \sum_{x\,\in\,\mathcal{{X}}}\pi\left(y\right)T\left(y,x\right)\\
 & = & \pi\left(y\right)\,\sum_{x\,\in\,\mathcal{{X}}}T\left(y,x\right)\\
 & = & \pi\left(y\right)\end{eqnarray*}

Remember that \emph{T} is stochastic.

\pagebreak

\end{document}